\documentclass[12pt]{article}             %%% twoside only for report
\usepackage{feynmp}                       %%% for Feynman diagrams; a few
\usepackage[dvips]{color}                 %%% for colours in the graphs
\usepackage{graphicx}                     %%% for pictures and figures
\usepackage{amssymb}

\newcommand{\comment}[1]{}
\newcommand{\note}[1]{}

\newcommand{\ia}{{\"{\i}}}   %not necessary if \usepackage[T1]{fontenc} is used
\newcommand{\absatz}{\vspace{2ex}\noindent}

\newcommand{\journal}[4]{{#1}\textbf{#2}, #3 (#4)}
\newcommand{\NPA}{\emph{Nucl.\ Phys.\ }\textbf{A}}
\newcommand{\NPB}{\emph{Nucl.\ Phys.\ }\textbf{B}}
\newcommand{\PLB}{\emph{Phys.\ Lett.\ }\textbf{B}}
\newcommand{\PR}{\emph{Phys.\ Rev.\ }}
\newcommand{\PRB}{\PR\textbf{B}}
\newcommand{\PRC}{\PR\textbf{C}}

\newcommand{\PRL}{\PR\emph{Lett.\ }}

\newcommand{\non}{\nonumber}

\newcommand{\half}{\frac{1}{2}}

\newcommand{\ii}{\mathrm{i}}

\newcommand{\T}{\mathrm{T}}

\newcommand{\kv}{\vec{k}}
\newcommand{\pv}{\vec{\,\!p}\!\:{}}
\newcommand{\qv}{\vec{\,\!q}\!\:{}}

\newcommand{\MeV}{\mathrm{MeV}}
\newcommand{\fm}{\mathrm{fm}}

\newcommand{\de}{\partial}
\newcommand{\dev}{\vec{\de}}

\newcommand{\calI}{\mathcal{I}}

\begin{document}

\begin{fmffile}{higfeyn}
  \fmfset{curly_len}{2mm}
  \fmfset{dash_len}{1.5mm}
  \fmfset{wiggly_len}{3mm}
  \newcommand{\feynbox}[2]{\mbox{\parbox{#1}{#2}}}
  \newcommand{\fs}{\scriptstyle}% adjusts the size of labels in feynmf-diagrams
  \newcommand{\hq}{\hspace{0.5em}}
  \newcommand{\hqm}{\hspace{-0.25em}}

\fmfcmd{vardef ellipseraw (expr p, ang) = save radx; numeric radx; radx=6/10
  length p; save rady; numeric rady; rady=3/10 length p; pair center;
  center:=point 1/2 length(p) of p; save t; transform t; t:=identity xscaled
  (2*radx*h) yscaled (2*rady*h) rotated (ang + angle direction length(p)/2 of
  p) shifted center; fullcircle transformed t enddef;
  style_def ellipse expr p= shadedraw ellipseraw (p,0); enddef;}
  
%%%%%%%%%%%%%%%%%%%%%%%%%%%%%%%%%%%%%%%%%%%%%%%%%%%%%%%%%%%%%%%%%%%%%%%%%%%%%%%
% This is a nice title page including abstract ....
%

\begin{titlepage}
\begin{flushright}
  nucl-th/9911034\\ DOE/ER/40561-74-INT99 \\ DUKE-TH-99-198
  \\ NT@UW-99-61 \\ TUM-T39-99-23 \\ 9th November 1999 \\
  Final, revised version\\
  12th April 2000\\
\end{flushright}
%\vspace*{1cm}
\begin{center}
  
  \LARGE{\textbf{Higher Partial Waves in an Effective Field Theory Approach to
      $nd$ Scattering}}

\end{center}
\vspace*{0.5cm}
\begin{center}
  
  \textbf{Fabrizio Gabbiani${}^{a, b, }$\footnote{Email: fg@phy.duke.edu;
      permanent address: (a)}},
  \textbf{Paulo F.\ Bedaque${}^{c,
      }$\footnote{Email: bedaque@phys.washington.edu}} and
  \textbf{Harald W.\ Grie\3hammer${}^{b, d, }$\footnote{Email:
      hgrie@physik.tu-muenchen.de; permanent address: (d)}}
  
  \vspace*{0.5cm}
  
  \emph{${}^a$Department of Physics, Duke University,\\
    Box 90305, Durham, NC 27708-0305, USA\\
    ${}^b$Nuclear Theory Group,
    Department of Physics, University of Washington,\\
    Box 351 560, Seattle, WA 98195-1560, USA\\
    ${}^c$Institute for Nuclear Theory,
    University of Washington,\\
    Box 351 550, Seattle, WA 98195-1550, USA\\
    ${}^d$Institut f{\"u}r Theoretische Physik, Physik-Department,\\
    Technische Universit{\"a}t M{\"u}nchen, 85747 Garching, Germany}
  \vspace*{0.2cm}

\end{center}

\vspace*{0.5cm}

\begin{abstract}
  The phase shifts for the higher partial waves ($l\ge 1$) in the spin quartet
  and doublet channel of $nd$ scattering at centre-of-mass energies up to
  $15\;\MeV$ are presented at next-to-leading and next-to-next-to-leading
  order in an effective field theory in which pions are integrated out. As
  available, the results agree with both phase shift analyses and potential
  model calculations.
\end{abstract}
\vskip 1.0cm
\noindent
\begin{tabular}{rl}
Suggested PACS numbers:& 11.80.J, 21.30.-x, 21.45.+v, 25.10.+s, 25.40.Dn\\[1ex]
Suggested Keywords: &\begin{minipage}[t]{10cm}
                    effective field theory, nucleon-deuteron
                    scattering, three-body systems
                    \end{minipage}
\end{tabular}
\end{titlepage}

\setcounter{page}{2} \setcounter{footnote}{0} \newpage

%%%%%%%%%%%%%%%%%%%%%%%%%%%%%%%%%%%%%%%%%%%%%%%%%%%%%%%%%%%%%%%%%%%%%%%%%%%%%%%
%%%%%%%%%%%%%%%%%%%%%%%%%%%%%%%%%%%%%%%%%%%%%%%%%%%%%%%%%%%%%%%%%%%%%%%%%%%%%%%
%%%%%%%%%%%%%%%%%%%%%%%%%%%%%%%%%%%%%%%%%%%%%%%%%%%%%%%%%%%%%%%%%%%%%%%%%%%%%%%
% Main Body
%

\section{Introduction}
\label{sec:intro}
\setcounter{equation}{0}

In the past few years, the Effective Field Theory
approach~\cite{LepageQEDlecture} has proven a useful tool to describe the
nuclear few body problem. Like the related low energy theories of QCD, Chiral
Perturbation Theory~\cite{leut,Weinberg79} and Heavy Baryon Chiral
Perturbation Theory~\cite{Gasseretal,JenkinsManohar} in the zero and one
nucleon sector respectively, it uses the fact that QCD exhibits a separation
of scales: The scales associated with confinement and chiral symmetry are so
high that it is natural to formulate QCD at low energies typical for nuclear
processes only in terms of the degrees of freedom which emerge after quarks
and gluons are confined in bound states. Below the pion cut, even pions are
absent and nucleons emerge as the only effective degrees of freedom. A low
energy expansion of QCD is then an expansion in powers of the typical momentum
of the process $Q$ in units of the high energy scale $\Lambda$ at which the
theory breaks down. This theory both with and without explicit pions was put
to extensive tests in the two-nucleon sector in recent
years~\cite{em}-\cite{iso}, starting with the pioneering work of Weinberg who
suggested the usefulness of EFT's in nuclear physics~\cite{Weinberg}. It
allowed for the first time for predictions which are systematic, rigorous and
model-independent (i.e.~independent of assumptions about the underlying
non-perturbative QCD dynamics). In most cases investigated, the experimental
agreement is within the estimated theoretical uncertainties, and in some
cases, previously unknown counterterms describing short distance physics could
be determined. Although in general process dependent, the expansion parameter
is found to be of the order of $\frac{1}{3}$ in most applications, so that NLO
calculations can be expected to be accurate to about $10\%$, and NNLO
calculations to about $4\%$. The theory with pions integrated out was recently
pushed to very high orders in the two-nucleon sector~\cite{CRS} where
accuracies of the order of $1\%$ were obtained. It can be viewed as a
systematisation of the Effective Range Theory (ERT) with the inclusion of
relativistic and short distance effects not found in traditional ERT
treatments.

The simplest problem in the three-nucleon sector is $nd$ scattering in the
spin-quartet $\mathrm{S}$ wave: The absence of Coulomb interactions ensures
that only properties of the strong interactions are probed, and the Pauli
principle guarantees that the three nucleons cannot occupy the same point in
space, making the effect of three-body forces very small.  The phase shifts in
this channel were shown to be well reproduced with and without pions both
below and above the deuteron breakup point up to momenta of $400\;\MeV$ in the
centre of mass frame~\cite{pbhg}. The explicit inclusion of pions was shown
not to improve the accuracy even at those relatively high momenta.  In the
$\mathrm{S}$ wave, spin-doublet channel (the triton channel) the situation is
more complicated. An unusual renormalisation of the three-body force makes it
large and as important as the leading two-body forces. Still, by using the
triton binding energy to fix the value of the three-body force, a reasonable
description of the phase shift in the doublet channel can be
obtained~\cite{Stooges2}.  In this article, we apply the EFT with pions
integrated out to the higher partial waves in the $nd$ system, above and below
the deuteron breakup point. All parameters are determined from $NN$
scattering, and three-body forces do not enter at the order we calculate,
allowing one to determine the range of validity of the pion-less EFT without a
detailed analysis of the fitting procedure. Since some problems depending
critically on some of these channels remain unsolved, like the $A_y$ puzzle,
the calculations presented are a testing ground for the applicability of the
EFT approach to these questions. 

\absatz In Sect.~\ref{sec:formalism}, we present a sketch of the EFT in the
many nucleon sector with pions integrated out. The application to $nd$
scattering in Sect.~\ref{sec:ndsystem} is followed by a discussion of our
results in the last Section.

%%%%%%%%%%%%%%%%%%%%%%%%%%%%%%%%%%%%%%%%%%%%%%%%%%%%%%%%%%%%%%%%%%%%%%%%%%%%%%%

\section{Formalism}
\label{sec:formalism}
\setcounter{equation}{0}

\subsection{Lagrangean and Power Counting}
\label{sec:lagrangean}

We now sketch the theory underlying our calculations, with more details to be
found in~\cite{pbhg}. Three main ingredients enter in the formulation of an
EFT: the Lagrangean, the power counting and a regularisation scheme.

After identifying the relevant degrees of freedom in a low energy theory of
QCD as the nucleons, one writes down the most general Lagrangean compliant
with the symmetries of QCD. Since a theory without explicit pions is expected
to break down approximately at the pion cut, $\Lambda\approx 100\; \MeV$, the
nucleons can be treated as non-relativistic particles with Lorentz invariance
restored by higher order terms. Without pions, only contact interactions are
allowed between nucleons. The first few terms in the most general, iso-spin
invariant Lagrangean are therefore
\begin{eqnarray}\label{ksw}
   \mathcal{L}_{NN}&=&N^\dagger(\ii
   \de_0+\frac{\dev^2}{2M}-\frac{\de_0^2}{2M})N- \non\\
   &&-\;C_{0\,d}(N^\T P^i_d N)^\dagger\  (N^\T P^i_dN) - \non\\
   &&
   -\;C_{0\,t} (N^\T P^A_t N)^\dagger\ (N^\T P^A_tN)+
   \\
   &&+ \;\frac{C_{2\,d}}{8}
   \left[(N^\T P^i_d N)^\dagger\ (N^\T P^i_d
     (\stackrel{\scriptscriptstyle\rightarrow}{\de}-
      \stackrel{\scriptscriptstyle\leftarrow}{\de})^2 N)\;+\;
   \mathrm{H.c.}\right]+
   \non\\
   &&
   + \;\frac{C_{2\,t}}{8}
   \left[(N^\T P^A_t N)^\dagger\ (N^\T P^A_t
     (\stackrel{\scriptscriptstyle\rightarrow}{\de}-
      \stackrel{\scriptscriptstyle\leftarrow}{\de})^2 N)\;+\;
   \mathrm{H.c.}\right]+\non
 \\
   &&+ \dots\;\;,\nonumber
\end{eqnarray}
where $N={p\choose n}$ is the nucleon doublet of two-component spinors and the
sub-scripts $d$ and $t$ denote the ${}^3\mathrm{S}_1$ and ${}^1\mathrm{S}_0$
channel of $NN$ scattering. For example, $P^i_d$ and $P^A_t$ are the
projectors onto the iso-scalar vector and scalar iso-vector channels,
\begin{equation}\label{proj}
  \left(P^i_d\right)^{b\beta}_{a\alpha}=
  \frac{1}{\sqrt{8}}\; (\sigma_2\sigma^i)_\alpha^\beta
  \;(\tau_2)_a^b \;\;,\;\;
  \left(P^A_t\right)^{b\beta}_{a\alpha}=
  \frac{1}{\sqrt{8}}\; (\sigma_2)_\alpha^\beta
  \;(\tau_2\tau^A)_a^b \;\;,
\end{equation}
with $\sigma$ ($\tau$) the Pauli matrices acting in spin (iso-spin) space. A
three-nucleon force appears at low orders only in the doublet $\mathrm{S}$
channel~\cite{Stooges2}.  The last of the nucleon kinetic energy terms in the
Lagrangean restores Lorentz invariance~\cite{CRS}. The coefficients $C_i$
encode all short distance physics -- like pion and $\omega$ exchanges, quarks
and gluons, and resonances like the $\Delta$ -- as the strengths of potentials
built out of derivatives of $\delta$ functions. In principle, these constants
could be derived by solving QCD or from models of short distance physics, but
the most common and practical way is to determine them from experiment.

\absatz Since the Lagrangean (\ref{ksw}) consists of infinitely many terms
only restricted by symmetry, predictive power is ensured only by the second
ingredient of an EFT: a power counting scheme, i.e.~a way to determine at
which order in a momentum expansion different contributions will appear,
keeping only and all the terms up to a given order in calculations. The
dimensionless, small parameter on which the expansion is based is the typical
momentum $Q$ of the process in units of the scale $\Lambda$ at which the
theory is expected to break down, e.g.~in a pion-less theory in units of the
pion mass. Assuming that all contributions are of natural size, i.e.\ ordered
by powers of $Q$, the systematic power counting ensures that the sum of all
terms left out when calculating to a certain order in $Q$ is smaller than the
last order retained. This way, an EFT allows for an error estimate of the
accuracy of the final result.

In the two-nucleon sector, finding such a power counting scheme is complicated
by the fact that the scattering lengths of both the ${}^1\mathrm{S}_0$ and
${}^3\mathrm{S}_1$ channel in $NN$ scattering are unnaturally large
($1/a_t=-8.3\;\MeV,\; 1/a_d=36\;\MeV$) because of the existence of a shallow
virtual and real bound state, the latter being the deuteron with a binding
energy $B=2.225\;\MeV$ and hence a typical binding momentum $\gamma_d=\sqrt{M
  B}\simeq 46\;\MeV$. This lies well below the expected breakdown scale of the
EFT and suggests that the system is close to a non-trivial fixed
point\footnote{See e.g.~\cite{BMR} for an analysis of this short distance fine
  tuning problem in terms of critical points of the renormalisation group.}.
Due to this fine tuning, the na{\ia}ve low momentum expansion has a very
limited range of validity ($Q\lesssim 1/a_t,Q\lesssim 1/a_d$). The alternative
is to expand in powers of the small parameter $Q/\Lambda \ll1$ but to keep the
full dependence on $Q a_t, Q a_d \sim 1$.  This is clearly necessary if one is
to be able to describe bound states within the effective theory. Some
interactions have therefore to be treated non-perturbatively in order to
accommodate these bound states. In the effective theory, the fine tuning
present in the two-body interactions arises as cancellation between loop and
counterterm contributions. This makes it tricky to do the power counting in a
na{\ia}ve approach since it cannot be applied diagram by diagram, but only to
whole classes of diagrams which are furthermore not always easily
identified~\cite{BiraAleph}. A convenient way of dealing with this problem was
suggested in~\cite{KSW}. It consists in using dimensional regularisation and a
new subtraction scheme (Power Divergence Subtraction, PDS) in which not only
the poles Feynman graphs exhibit in $4$ dimensions are subtracted but also the
poles in $3$ dimensions. This regularisation procedure is chosen to explicitly
preserve the systematic power counting as well as all symmetries at each order
in every step of the calculation. If one chooses the arbitrary scale $\mu$
arising in the subtraction to be of the order $\mu \sim Q\sim 1/a_t\sim
1/a_d$, one finds
\begin{eqnarray}\label{scalingksw}
  C_0^{(-1)}&\sim&\frac{1}{M Q}\;\;,\nonumber\\
  C_0^{(0)}&\sim&\frac{1}{M \Lambda}\;\;,\\
  C_2&\sim&\frac{1}{M \Lambda Q^2}\nonumber
\end{eqnarray}
for the short distance coefficients of both the ${}^3\mathrm{S}_1$ and
${}^1\mathrm{S}_0$ channel. Here, as in the following, we split the
coefficients of the leading four point interactions, $C_0$, into a leading
($C_0^{(-1)}$) and a sub-leading piece ($C_0^{(0)}$), where the super-script
in parenthesis denotes the scaling of the coefficient with $Q$.

\absatz Using (\ref{scalingksw}) one can estimate the contribution of any
given diagram.  As will be demonstrated in the next sub-section, it is found
that for nucleon-nucleon scattering, the leading order contribution is given
by an infinite number of diagrams forming a bubble chain, with the vertices
being the ones proportional to $C_0$. Higher order corrections are
perturbative and given by one or more insertions of higher derivative
operators.  In the three-body sector discussed in this paper, even the leading
order calculation is too complex for a fully analytical solution.  Still, the
equations that need to be solved are computationally trivial and can
furthermore be improved systematically by higher order corrections that
involve only integrations, as opposed to many-dimensional integral equations
arising in other approaches.

\absatz Following a previous paper on quartet $\mathrm{S}$ wave scattering in
the $nd$ system~\cite{pbhg}, we choose a Lagrangean which is equivalent to
(\ref{ksw}) but contains two auxiliary fields $d^i$ and $t^A$ with the quantum
numbers of the deuteron and of a di-baryon field in the ${}^1\mathrm{S}_0$
channel of $NN$ scattering respectively, such that the four nucleon
interactions are removed:
\begin{eqnarray}\label{dlag}
   \mathcal{L}_{Nd}&=&N^\dagger (\ii \partial_0 +\frac{\dev^2}{2
     M}-\frac{\de_0^2}{2M})N+\non\\
   &&-d^{i  \dagger} \left[(\ii \partial_0
     +\frac{\dev^2}{4M})+\Delta^{(-1)}_d+\Delta^{(0)}_d\right]d^i
       +\;y_d\left[d^{i \dagger} (N^\T P^i_dN) \;+\;\mathrm{H.c.}\right] +\\
   &&-t^{A \dagger} \left[(\ii \partial_0 +\frac{\dev^2}{4
       M})+\Delta^{(-1)}_t+\Delta^{(0)}_t\right]t^A
   +\;y_t\left[t^{A \dagger} (N^\T P^A_t
     N) \;+\;\mathrm{H.c.}\right] +
   \dots\non
\end{eqnarray}
It was demonstrated in~\cite{pbhg} that this Lagrangean is more convenient for
the numerical investigations necessary in the three-body system.  Analogously
to the $C_0$'s, we split the $\Delta$'s into leading ($\Delta^{(-1)}$) and
sub-leading pieces ($\Delta^{(0)}$). Performing the Gau\3ian integration over
the auxiliary fields $d^i$ and $t^A$ in the path integral followed by a field
re-definition in order to eliminate the terms containing time derivatives, we
see that the two Lagrangeans (\ref{ksw}) and (\ref{dlag}) are indeed
equivalent up to higher order terms when one identifies
\begin{eqnarray}\label{ytoC}
  \Delta^{(-1)}&=&-\frac{C_0^{(-1)}}{M C_2}\;\;,\nonumber\\
  \Delta^{(0)}&=&\frac{C_0^{(0)}}{M C_2}\;\;,\\
  y^2&=&\frac{(C_0^{(-1)})^2}{M C_2}\;\;\nonumber
\end{eqnarray}
for each of the two $\mathrm{S}$ wave channels separately. The ``wrong'' sign
of the kinetic energy terms of the auxiliary fields therefore does not spoil
unitarity. The new coefficients scale from (\ref{scalingksw}) and (\ref{ytoC})
in each channel as
\begin{eqnarray}\label{deltascaling}
\Delta^{(-1)}&\sim&\frac{Q \Lambda}{M}\;\;,\nonumber\\
 \Delta^{(0)}&\sim&\frac{Q^2}{M}\;\;,\\
 y^2&\sim&\frac{\Lambda}{M^2}\;\;.\nonumber
\end{eqnarray}

\subsection{The Two-Nucleon System to NNLO}
\label{sec:twonucleons}

It is useful to consider the solution to the two-body problem before turning
to the three-body problem. As the ${}^3\mathrm{S}_1$ and ${}^1\mathrm{S}_0$
channels do not differ in their power counting, we concentrate on the deuteron
channel. The considerations in the ${}^1\mathrm{S}_0$ channel are analogous.
More details about the EFT without pions in the triplet channel can be found
in~\cite{CRS}.

\begin{figure}[!htb]
  \begin{center}
    \feynbox{40\unitlength}{
            \begin{fmfgraph*}(40,40)
              \fmfleft{i} \fmfright{o} \fmf{double,width=thin}{i,o}
            \end{fmfgraph*}}
          \hq$=$\hq \feynbox{40\unitlength}{
            \begin{fmfgraph*}(40,40)
              \fmfleft{i} \fmfright{o} \fmf{vanilla,width=1.5*thick}{i,o}
            \end{fmfgraph*}}
          \hq$+$\hq \feynbox{40\unitlength}{
            \begin{fmfgraph*}(40,40)
              \fmfleft{i} \fmfright{o}
              \fmf{vanilla,width=1.5*thick,tension=5}{i,v1}
              \fmf{vanilla,width=1.5*thick,tension=5}{o,v2}
              \fmf{vanilla,width=thin,left=0.65}{v1,v2}
              \fmf{vanilla,width=thin,left=0.65}{v2,v1}
            \end{fmfgraph*}}
          \hq$+$\hq \feynbox{70\unitlength}{
            \begin{fmfgraph*}(70,40)
              \fmfleft{i} \fmfright{o}
              \fmf{vanilla,width=1.5*thick,tension=5}{i,v1}
              \fmf{vanilla,width=1.5*thick,tension=5}{v2,v3}
              \fmf{vanilla,width=1.5*thick,tension=5}{v4,o}
              \fmf{vanilla,width=thin,left=0.65}{v1,v2}
              \fmf{vanilla,width=thin,left=0.65}{v2,v1}
              \fmf{vanilla,width=thin,left=0.65}{v3,v4}
              \fmf{vanilla,width=thin,left=0.65}{v4,v3}
            \end{fmfgraph*}}
          \hq$+$\hq \feynbox{100\unitlength}{
            \begin{fmfgraph*}(100,40)
              \fmfleft{i} \fmfright{o}
              \fmf{vanilla,width=1.5*thick,tension=5}{i,v1}
              \fmf{vanilla,width=1.5*thick,tension=5}{v2,v3}
              \fmf{vanilla,width=1.5*thick,tension=5}{v4,v5}
              \fmf{vanilla,width=1.5*thick,tension=5}{v6,o}
              \fmf{vanilla,width=thin,left=0.65}{v1,v2}
              \fmf{vanilla,width=thin,left=0.65}{v2,v1}
              \fmf{vanilla,width=thin,left=0.65}{v3,v4}
              \fmf{vanilla,width=thin,left=0.65}{v4,v3}
              \fmf{vanilla,width=thin,left=0.65}{v5,v6}
              \fmf{vanilla,width=thin,left=0.65}{v6,v5}
            \end{fmfgraph*}}
          \hq$+\;\dots$
  \end{center}
  \caption{\label{fig:deuteronpropagator} \sl The deuteron propagator
    at leading order from the Lagrangean (\protect\ref{dlag}). The thick solid
    line denotes the bare propagator \protect$\frac{-\ii}{\Delta^{(-1)}_d}$,
    the double line its dressed counterpart.}
\end{figure}
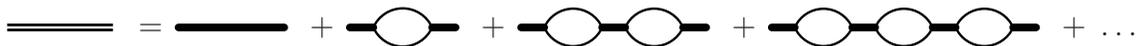

Any diagram can be estimated by scaling all momenta by a factor of $Q$ and all
energies by a factor of $Q^2/M$.  The remaining integral includes no
dimensions and is taken to be of the order $Q^0$ and of natural size. This
scaling implies the rule that nucleon propagators contribute one power of
$M/Q^2$ and each loop a power of $Q^5/M$.  Relativistic corrections to the
kinetic energy of the nucleon scale like $\frac{Q^4}{M^3}$ and hence only
enter as insertions into the non-relativistic nucleon propagator at NNLO in
$Q$, suppressed by additional powers of $1/M$. The vertices provide powers of
$Q$ according to (\ref{deltascaling}), implying that the deuteron kinetic
energy term is sub-leading compared to the $\Delta^{(-1)}_d$ term.  Thus, the
bare deuteron propagator is just the constant $-\ii/\Delta^{(-1)}_d$.  Using
these rules, there is an infinite number of diagrams contributing at leading
order to the deuteron propagator, as shown in
Fig.~\ref{fig:deuteronpropagator}, each one of the order $1/(MQ)$.
The linear divergence in each of the bubble diagrams shown in
Fig.~\ref{fig:deuteronpropagator} does not show in dimensional regularisation
as a pole in $4$ dimensions, but it does appear as a pole in $3$ dimensions
which we subtract following the PDS scheme.

The full leading order propagator $\ii\triangle^{ij}_d(p)$ of the deuteron
field consists hence of the geometric series shown in
Fig.~\ref{fig:deuteronpropagator},
\begin{equation}\label{dprop}
  \ii\triangle^{ij}_d(p) =-\;\frac{4\pi \ii}{M y^2_d}\;
  \frac{\delta^{ij}}{\frac{4\pi
  \Delta^{(-1)}_d}{M y^2_d} -\mu+\sqrt{\frac{\pv^2}{4}-M p_0-\ii\varepsilon}}
  \;\;,
\end{equation}
where $\mu$ is the arbitrary renormalisation scale introduced by the PDS
scheme, showing that $\Delta^{(-1)}_d/y^2_d$ is renormalisation group
dependent, too. Physical observables like amplitudes are of course independent
of the choice of $\mu$, as will be demonstrated now. We determine the free
parameters by demanding that the deuteron propagator has its pole at the
physical deuteron pole position as
\begin{equation}\label{fitLO}
  -\frac{y^2_d}{\Delta^{(-1)}_d}=\frac{4
  \pi}{M}\;\frac{1}{\gamma_d-\mu}=C_{0\,d}^{(-1)}\;\;.
\end{equation}
Since $\gamma_d\sim Q$, the choice $\mu\sim Q$ makes the coefficient
$C_{0\,d}^{(-1)}$ indeed scale as required in the power counting scheme
(\ref{scalingksw}).  The deuteron and di-baryon propagators are hence in terms
of physical quantities at leading order given by
\begin{eqnarray}\label{props}
  \ii\triangle^{ij}_{d}(p) \;=\;\ii\delta^{ij}\,\triangle_{d}(p)& =&
  \frac{4\pi \ii}{M y^2_{d}}\;
  \frac{\delta^{ij}}{\gamma_{d}-\sqrt{\frac{\pv^2}{4}-M
  p_0-\ii\varepsilon}}\;\;,\non\\
  \ii\triangle^{AB}_{t}(p) \;=\;\ii\delta^{AB}\,\triangle_{t}(p)& =&
  \frac{4\pi \ii}{M y^2_{t}}\;
  \frac{\delta^{AB}}{\gamma_{t}-\sqrt{\frac{\pv^2}{4}-M
  p_0-\ii\varepsilon}}\;\;.
\end{eqnarray}
The typical momentum $\gamma_t=1/a_t$ of the virtual bound state is extracted
from the scattering length in this channel. Each deuteron or di-baryon
propagator is accompanied by a power of $1/(M y^2 Q)= M/(\Lambda Q)$.

The kinetic energy insertion and the mass insertion proportional to
$\Delta^{(0)}_d$ are suppressed by one power of $Q$ according to the scaling
properties (\ref{deltascaling}) and hence enter at NLO.  We choose to keep the
$\Delta^{(-1)}$'s to be the same as at LO, so that (\ref{fitLO}) is still
valid at NLO and higher order calculations will not necessitate a re-fitting
of lower order coefficients~\cite{RupakShoresh}. The remaining pieces,
$\Delta^{(0)}$, are then parametrically smaller (as indicated by
(\ref{deltascaling})) and are included perturbatively. Two conditions per
partial wave are now necessary to fix the new constants at NLO. One condition
is that the pole position of the bound state does not change,
Fig.~\ref{fig:polecondition}:
\begin{figure}[!htb]
  
  \vspace*{3ex}
  
  \begin{center}
    $ \left[ \feynbox{40\unitlength}{
            \begin{fmfgraph*}(40,20)
              \fmfleft{i} \fmfright{o} \fmf{double,width=thin}{i,v1,o}
              \fmfblob{8*thick}{v1}
            \end{fmfgraph*}}
        \right]^{-1}\bigg|_{p_0=-\frac{\gamma^2_d}{M},\,\pv=0}\stackrel{!}{=}0$
        \hq:\hq\hq\\[3ex]
        $ \left[ \feynbox{40\unitlength}{
            \begin{fmfgraph*}(40,20)
              \fmfleft{i} \fmfright{o} \fmf{double,width=thin}{i,o}
            \end{fmfgraph*}}
        \right]^{-1}\bigg|_{p_0=-\frac{\gamma^2_d}{M},\atop\,\pv=0}
        \stackrel{!}{=}0$ \hq\hq,\hq\hq $ \left[\hq \feynbox{40\unitlength}{
            \begin{fmfgraph*}(40,40)
              \fmfleft{i} \fmfright{o} \fmf{double,width=thin}{i,v1,o}
              \fmfv{decoration.shape=cross,decor.size=6*thick,label=$\fs-\ii
                (p_0-\frac{\pv^2}{4M})$,label.angle=90}{v1}
            \end{fmfgraph*}}
          \hq+\hq \feynbox{40\unitlength}{
            \begin{fmfgraph*}(40,40)
              \fmfleft{i} \fmfright{o} \fmf{double,width=thin}{i,v1,o}
              \fmfv{decoration.shape=circle,decor.filled=empty,
                decor.size=3*thick,
                label=$\fs-\ii\Delta^{(0)}_d$,label.angle=90}{v1}
            \end{fmfgraph*}}
        \right]_{p_0=-\frac{\gamma^2_d}{M},\atop\,\pv=0}\stackrel{!}{=}0$
  \end{center}
  \caption{\label{fig:polecondition} \sl The first condition on the
    coefficients \protect$\Delta^{(-1)}_d,\,\Delta^{(0)}_d$ and \protect$y_d$,
    (\protect\ref{fitLO}/\protect\ref{polecondition}). The dressed deuteron
    propagator has its pole at the physical binding energy at LO, and higher
    order corrections do not change its position. The cross denotes a kinetic
    energy insertion, the dot an insertion of \protect$\Delta^{(0)}_d$.}
\end{figure}
\begin{equation}
  \label{polecondition}
  \Delta^{(0)}_d=\frac{\gamma_d^2}{M}
\end{equation}
The second condition in the ${}^3\mathrm{S}_1$ channel can be that the
deuteron pole has the correct residue $Z_d$~\cite{PRS},
\begin{equation}
  \label{residue}
  \frac{8\pi\gamma_d}{M^2 y^2_d}=Z_d-1\;\;,\;\;Z_d^{-1}=1-\gamma_d\rho_d\;\;,
\end{equation}
where $\rho_d$ is defined by the effective range expansion around the deuteron
pole,
\begin{equation}
   k  \cot \delta=-\gamma_d+\frac{1}{2} \rho_d ( k^2+\gamma^2_d) + \dots\;\;,
\end{equation}
for the nucleon-nucleon phase shifts.  In the singlet channel, no real bound
state exists whose properties have to be described, so we impose the condition
that the effective range expansion
\begin{equation}
  \label{acondition}
   k \cot\delta=-{1 \over a_t}+\frac{1}{2} r_{0t} k^2 + \dots
\end{equation}
is reproduced.  Solving (\ref{polecondition}/\ref{residue}/\ref{acondition})
with (\ref{fitLO}) leads to
\begin{eqnarray}
  \label{fitNLO}
  \Delta^{(-1)}_d= \frac{2}{M}\;\frac{\mu-\gamma_d}{\rho_d}\ &,&
  \Delta^{(-1)}_t= \frac{2}{M}\;\frac{\mu-\gamma_t}{r_{0t}} \;\;,\non\\
  \Delta^{(0)}_d=\frac{\gamma^2_d}{M} &,&
  \Delta^{(0)}_t= 0\;\;,\\
  y^2_d= \frac{8\pi}{M^2}\;\frac{\gamma_d}{Z_d-1}&,&
  y^2_t= \frac{8 \pi}{M^2}\;\frac{1}{r_{0t}}\;\;.\non
\end{eqnarray}
In a NNLO calculation, $C_2$ is split like $C_0$, with $C_2^{(-2)}$ still
given by the expression above, $C_2^{(-1)}$ a NNLO correction parametrically
small against $C_2^{(-2)}$. Likewise, $C_0$ has now a NNLO component
$C_0^{(1)}$ which is again fixed by the condition that the deuteron or
di-baryon pole position is unchanged. This way, higher order calculations do
not necessitate a re-fitting of lower order coefficients~\cite{RupakShoresh}.
In addition, four new interactions enter at NNLO~\cite{CRS}: The first is a
term which induces $\mathrm{SD}$ mixing in the two-nucleon system and will be
shown to be irrelevant for our problem.  Relativistic corrections to the
nucleon propagator, $-\frac{\de_0^2}{2M}\sim\frac{Q^4}{M^3}$, enter at NNLO in
the $Q$ counting, but are suppressed by two more powers of $M$ and hence can
be neglected.  Finally, there are $NN$ interactions containing four
derivatives. In the triplet channel, for example, a term
\begin{equation}
  \label{eq:c4}
  -\frac{1}{16}\; C_{4\,d}\;
   (N^\T P^i_d
     (\stackrel{\scriptscriptstyle\rightarrow}{\de}-
      \stackrel{\scriptscriptstyle\leftarrow}{\de})^2 N)^\dagger
 \;(N^\T P^i_d
     (\stackrel{\scriptscriptstyle\rightarrow}{\de}-
      \stackrel{\scriptscriptstyle\leftarrow}{\de})^2 N)
\end{equation}
enters in the Lagrangean (\ref{ksw}) which by Gau\3ian integration and field
re-definition can be shown to be equivalent to a term in the modified
Lagrangean (\ref{dlag}) of the form
\begin{equation}
  \label{eq:c4indlag}
  \frac{MC_{4\,d}}{C_{2\,d}^{(-2)}}\;d^{i
      \dagger}\left[\ii\de_0+\frac{\dev^2}{4M}\right]^2 d^i\;\;.
\end{equation}
However, it was shown in~\cite{CRS} that the renormalisation group equations
determine the coefficient $C_4$ through the lower order constants $C_0$ and
$C_2$, i.e.~that the value of this constant is determined already from the NLO
parameters as~\cite{PRS}
\begin{equation}
  \label{eq:c4determination}
  C_{4\,d}=-\frac{\pi}{M}\;\frac{(Z_d-1)^2}{\gamma^2_d(\mu-\gamma_d)^3}\;\;.
\end{equation}
This is a generic feature, the leading piece of these higher derivative terms
are determined by the LO and NLO coefficients. Their independent contribution
start at N3LO but can be re-summed into a full deuteron (di-baryon) propagator
\begin{eqnarray}
  \label{n3loprops}
  \ii\triangle^{ij}_{d}(p;\mathrm{N3LO}) &=&\ii\delta^{ij}\,
  \triangle_{d}(p;\mathrm{N3LO})
  \;=\non\\
  &=&-\;\frac{4\pi \ii}{M y^2_d}\;
  \frac{\delta^{ij}}{-\gamma_{d}+\frac{1}{2} \rho_d (M p_0-\frac{\pv^2}{4}
    +\gamma^2_d)+
  \sqrt{\frac{\pv^2}{4}-M p_0-\ii\varepsilon}}\;\;,\non\\
  \ii\triangle^{AB}_{t}(p;\mathrm{N3LO}) &=&\ii\delta^{AB}\,
  \triangle_{t}(p;\mathrm{N3LO})
  \;=\non\\
  &=&-\;\frac{4\pi \ii}{M y^2_t}\;
  \frac{\delta^{AB}}{-\gamma_t+\frac{1}{2} r_{0t} (M p_0-\frac{\pv^2}{4})+
  \sqrt{\frac{\pv^2}{4}-M p_0-\ii\varepsilon}}\;\;.
\end{eqnarray}

%%%%%%%%%%%%%%%%%%%%%%%%%%%%%%%%%%%%%%%%%%%%%%%%%%%%%%%%%%%%%%%%%%%%%%%%%%%%%%%

\section{The $nd$ System}
\label{sec:ndsystem}
\setcounter{equation}{0}

\subsection{Quartet Channel}
\label{sec:ndquartet}

Turning to the $nd$ system, we first consider the quartet channel as its
treatment is analogous to the doublet channel without the difficulties of a
coupled channel equation. For details, we refer again to the treatment of the
quartet $\mathrm{S}$ wave in~\cite{pbhg}.

\begin{figure}[!htb]
  \begin{center}
    
    \vspace*{3ex}
    
    \setlength{\unitlength}{0.8pt}
    
    \feynbox{104\unitlength}{
            \begin{fmfgraph*}(104,64)
              \fmfleft{i2,i1} \fmfright{o2,o1}
              \fmfv{label=$\fs(\frac{\kv^2}{4M}-\frac{\gamma^2_d}{M},,\kv)$,
                label.angle=90}{i1}
              \fmfv{label=$\fs(\frac{\kv^2}{4M}-\frac{\gamma^2_d}{M}+\epsilon
                ,,\pv)$,label.angle=90}{o1}
              \fmfv{label=$\fs(\frac{\kv^2}{2M},,-\kv)$,label.angle=-90}{i2}
              \fmfv{label=$\fs(\frac{\kv^2}{2M}-\epsilon,,-\pv)$,
                label.angle=-90}{o2} \fmf{double,width=thin,tension=3}{i1,v1}
              \fmf{double,width=thin,tension=1.5}{v1,v3,v2}
              \fmf{double,width=thin,tension=3}{v2,o1}
              \fmf{vanilla,width=thin}{i2,v4,o2} \fmffreeze
              \fmf{ellipse,rubout=1,label=$\fs t_{0\,,\mathrm{q}}^{ij}$,
                label.dist=0.25w,label.side=right}{v3,v4}
              \end{fmfgraph*}}
            \hq$=$\hq \feynbox{104\unitlength}{
            \begin{fmfgraph*}(104,64)
              \fmfleft{i2,i1} \fmfright{o2,o1}
              \fmf{double,width=thin,tension=4}{i1,v1,v2}
              \fmf{vanilla,width=thin}{v2,o1}
              \fmf{double,width=thin,tension=4}{v3,v4,o2}
              \fmf{vanilla,width=thin}{i2,v3} \fmffreeze
              \fmf{vanilla,width=thin}{v2,v3}
              \end{fmfgraph*}}
            \hq$+$\hq \feynbox{192\unitlength}{
            \begin{fmfgraph*}(192,64)
              \fmfleft{i2,i1} \fmfright{o2,o1}
              \fmf{double,width=thin,tension=3}{i1,v1}
              \fmf{double,width=thin,tension=1.5}{v1,v6,v5}
              \fmf{double,width=thin,tension=3}{v5,v2}
              \fmf{vanilla,width=thin}{v2,o1} \fmf{vanilla,width=thin}{i2,v7}
              \fmf{vanilla,width=thin,tension=0.666}{v7,v4}
              \fmf{double,width=thin,tension=4}{v4,v3,o2} \fmffreeze
              \fmffreeze \fmf{vanilla,width=thin}{v4,v2}
              \fmf{ellipse,rubout=1,label=$\fs t_{0\,,\mathrm{q}}^{ij}$,
                label.dist=0.15w,label.side=right}{v6,v7}
              \end{fmfgraph*}}
            
            \vspace*{2.5ex}
            
            \setlength{\unitlength}{1pt}
    
  \end{center}
    \caption{\label{fig:LOfaddeevquartet} \sl The Faddeev equation
      (\ref{faddeevequationquartet}) which needs to be solved for
      \protect$t_{0\,\mathrm{q}}^{ij}$ at LO in the quartet channel.}
\end{figure}
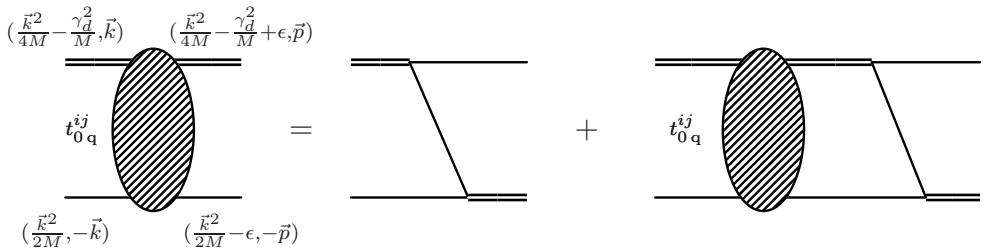

All three nucleon spins are aligned, so that only properties of the triplet
$NN$ system are probed.  It has been shown in~\cite{pbhg} that all diagrams in
which three nucleons interact only via the LO deuteron-nucleon potential $y_d$
are of the same, leading order $\Lambda/(MQ^2)$. Summing all ``bubble-chain''
sub-graphs of Fig.~\ref{fig:deuteronpropagator} into the deuteron propagator,
one arrives at the integral equation pictorially represented in
Fig.~\ref{fig:LOfaddeevquartet}.
To derive this equation for the half off-shell amplitude at LO, let us choose
the kinematics as in Fig.~\ref{fig:LOfaddeevquartet} in such a way that the
incoming (outgoing) deuteron line carries momentum $\kv$ ($\pv$), energy $
\kv^2/(4M) - \gamma^2_d/M$ ($\kv^2/(4M) - \gamma^2_d/M+\epsilon$) and vector
index $j$ ($i$).  The incoming (outgoing) nucleon line carries momentum $-\kv$
($-\pv$), energy $\kv^2/(2M)$ ($\kv^2/(2M) - \epsilon$) and spinor and
iso-spinor indices $\alpha$ and $a$ ($\beta$ and $b$). Hence, the incoming
deuteron and nucleon are on shell, and $\epsilon$ denotes how far the outgoing
particles are off shell. We denote by $\left(\ii
  t_{0\,\mathrm{q}}^{ij}\right)_{\alpha a}^{\beta b}(\vec{k},\pv,\epsilon)$
the sum of those diagrams with the kinematics above forming the half off-shell
amplitude in the quartet channel, and read off from the lower line of
Fig.~\ref{fig:LOfaddeevquartet} the integral equation:
\begin{eqnarray}
  \label{faddeevequationquartet}
  \left(\ii t_{0\,\mathrm{q}}^{ij}\right)_{\alpha a}^{\beta b}
  (\kv,\pv,\epsilon) &=& \frac{y^2_d}{2}\;
  (\sigma^j\sigma^i)_\alpha^\beta \;\delta_a^b\;
  \frac{\ii}{-\frac{\kv^2}{4M}-\frac{\gamma^2_d}{M}+\epsilon
    -\frac{(\kv+\pv)^2}{2M}+ \ii\varepsilon}+\non\\
  &  &+\;\frac{y^2_d}{2}\;
  (\sigma^j\sigma^k)_\gamma^\beta \;\delta_c^b\; \int
  \frac{d^4q}{(2\pi)^4}\
  \left(\ii t_{0\,\mathrm{q}}^{ik}\right)_{\alpha a}^{\gamma
    c}(\kv,\qv,\epsilon+q_0)\;\times\\
  & &\phantom{4y^2_d (\sigma^j\sigma^k)_\gamma^\beta}\times\;
  \ii\triangle_d(\frac{\kv^2}{4M}-\frac{\gamma^2_d}{M}+\epsilon+q_0,\qv)\;
  \frac{\ii}{\frac{\kv^2}{2M}-\epsilon-q_0-\frac{\qv^2}{2M}+ \ii\varepsilon}
  \times
  \nonumber\\
  & &\phantom{4y^2_d (\sigma^j\sigma^k)_\gamma^\beta}\times\;
  \frac{\ii}{-\frac{\kv^2}{4M}-\frac{\gamma^2_d}{M}+2\epsilon
    +q_0-\frac{(\qv+\pv)^2}{2M}+\ii\varepsilon}\;\;.\nonumber
\end{eqnarray}
The integration over $q_0$ picks the pole at $q_0=(\kv^2-\qv^2)/(2M)-\epsilon+
\ii\varepsilon$. After that, we set $\epsilon=(\kv^2-\pv^2)/(2M)$, integrate
over the angle between $\kv$ and $\pv$ weighted with the Legendre polynomial
of the first kind $P_l(\hat{\kv} \cdot \hat{\pv})$ to project onto the $l$th
partial wave and set $i=(1+\ii 2)/\sqrt{2},\;j=(1-\ii
2)/\sqrt{2},\;\alpha=\beta=1,\;a=b=1$, to pick up the spin quartet part.
Denoting this projected amplitude by $\ii t^{(l)}_{0\,\mathrm{q}}(k,p)$, we
find
\begin{eqnarray}\label{eqfort}
  t^{(l)}_{0\,\mathrm{q}}(k,p)&=&-\;\frac{y^2_d M}{p k}\;Q_l\left({{p^2+k^2-ME
- \ii\varepsilon} \over {pk}}\right)- \\
 &&-\;\frac{2}{\pi}\int\limits_0^\infty dq\; q^2\; t^{(l)}_{0\,\mathrm{q}}
 (k,q)\;
 \frac{1}{\sqrt{\frac{3 q^2}{4}-ME- \ii\varepsilon}-\gamma_d}\;\frac{1}{qp}\;
Q_l\left({{p^2 + q^2-ME  - \ii\varepsilon} \over {pq}}\right),\nonumber
\end{eqnarray}
where $E=3 k^2/(4M) - \gamma^2_d/M$ is the total energy and the Legendre
polynomials of the second kind $Q_l$ are defined as
\begin{equation}
Q_l(a) = \half\;\int\limits^{1}_{-1}\; dx\; {P_l(x) \over {x+a}}\;\;. 
\end{equation}
This Faddeev equation was first derived in~\cite{Skornyakov}. Notice that
when $p=k$, all external legs are on-shell. Although the pole on the real axis
due to the deuteron propagator is regulated by the $\ii\varepsilon$
prescription and the logarithmic singularities occurring above threshold are
integrable for each partial wave, both cause numerical instabilities. We used
the Hetherington--Schick~\cite{HetheringtonSchick,Amado,Book} method to
numerically solve (\ref{eqfort}). The basic idea is to perform a rotation of
the variable $q$ into the complex plane by an angle large enough in order to
avoid the singularities in and near the real axis but small enough not to
cross the singularities of the kernel or of the solution. One can show that
the singularities of the solution are not closer to the real axis than those
of the inhomogeneous, Born term~\cite{Brayshaw}. The solution on the real axis
can then be obtained from the solution on the deformed contour by using
(\ref{eqfort}) again, now with $k$ and $p$ on the real axis and $q$ on the
contour. The computational effort becomes trivial, and a code runs within
seconds on a personal computer.

To obtain the neutron-deuteron scattering amplitude, we have to multiply the
on-shell amplitude $t^{(l)}_{0\,\mathrm{q}}(k,p)$ by the wave function
renormalisation constant,
\begin{eqnarray}
  \label{Z0}
  T^{(l)}_{0\,\mathrm{q}}(k)&=&\sqrt{Z_0}\; t^{(l)}_{0\,\mathrm{q}}(k,k)\;
  \sqrt{Z_0}\;\; , \nonumber\\
 \frac{1}{Z_0}&=& \ii \frac{\partial}{\partial p_0}\;
 \frac{1}{\ii\triangle_d(p)} \Big|_{p_0=-\frac{\gamma^2_d}{M},\,\pv=0}
 \nonumber\\
 &= & \frac{M^2 y^2_d}{8\pi \gamma_d}\;\;.
\end{eqnarray}
In contradistinction to $\ii t^{(l)}_{0\,\mathrm{q}}(k,p)$, the scattering
amplitude $\ii T^{(l)}_{0\,\mathrm{q}}(k)$ depends on the parameters $y_d$ and
$\Delta^{(-1)}_d$ only through the observable $\gamma_d$.

\begin{figure}[!htb]
  \begin{center}
    
    \vspace*{5ex}
    
    \setlength{\unitlength}{0.75pt}

%Insertions
    
    \feynbox{200\unitlength}{
            \begin{fmfgraph*}(200,64)
              \fmfleft{i2,i1} \fmfright{o2,o1}
              \fmf{double,width=thin,tension=3}{i1,v1}
              \fmf{double,width=thin,tension=1.5}{v1,v3,v2}
              \fmf{double,width=thin,tension=3}{v2,v5}
              \fmfv{decor.shape=cross,decor.size=6*thick}{v5}
              \fmf{double,width=thin,tension=3}{v5,v7}
              \fmf{double,width=thin,tension=1.5}{v7,v8,v9}
              \fmf{double,width=thin,tension=3}{v9,o1}
              \fmf{vanilla,width=thin}{i2,v4}
              \fmf{vanilla,width=thin,tension=0.5}{v4,v10}
              \fmf{vanilla,width=thin}{v10,o2} \fmffreeze
              \fmf{ellipse,rubout=1,label=$\fs t_{0\,,\mathrm{q}}^{ij}$,
                label.dist=0.12w,label.side=right}{v3,v4}
              \fmf{ellipse,rubout=1,label=$\fs t_{0\,,\mathrm{q}}^{ij}$,
                label.dist=0.12w,label.side=left}{v8,v10}
              \fmf{vanilla,width=thin}{v3,v4} \fmf{vanilla,width=thin}{v8,v10}
              \end{fmfgraph*}}
            \hqm\hqm$+$\hqm\hqm \feynbox{200\unitlength}{
            \begin{fmfgraph*}(200,64)
              \fmfleft{i2,i1} \fmfright{o2,o1}
              \fmf{double,width=thin,tension=3}{i1,v1}
              \fmf{double,width=thin,tension=1.5}{v1,v3,v2}
              \fmf{double,width=thin,tension=3}{v2,v5}
              \fmfv{decor.shape=circle,decor.filled=empty, decor.size=3*thick,
                label=$\fs-\ii\Delta^{(0)}_d$,label.angle=90}{v5}
              \fmf{double,width=thin,tension=3}{v5,v7}
              \fmf{double,width=thin,tension=1.5}{v7,v8,v9}
              \fmf{double,width=thin,tension=3}{v9,o1}
              \fmf{vanilla,width=thin}{i2,v4}
              \fmf{vanilla,width=thin,tension=0.5}{v4,v10}
              \fmf{vanilla,width=thin}{v10,o2} \fmffreeze \fmffreeze
              \fmf{ellipse,rubout=1,label=$\fs t_{0\,,\mathrm{q}}^{ij}$,
                label.dist=0.12w,label.side=right}{v3,v4}
              \fmf{ellipse,rubout=1,label=$\fs t_{0\,,\mathrm{q}}^{ij}$,
                label.dist=0.12w,label.side=left}{v8,v10}
              \fmf{vanilla,width=thin}{v3,v4} \fmf{vanilla,width=thin}{v8,v10}
              \end{fmfgraph*}}
  \end{center}
    \caption{\label{fig:NLOthreebodyquartet} \sl The NLO contributions to $nd$
      scattering in the quartet channel. Notation as in
      Fig.~\protect\ref{fig:polecondition}.}
\end{figure}
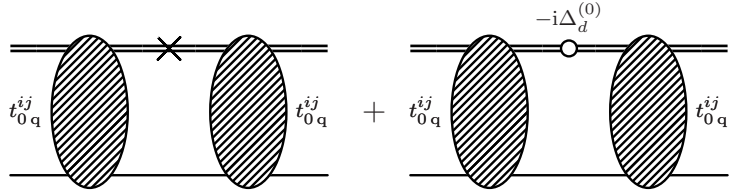

\absatz At NLO, we have additional contributions only from modifications of
the deuteron, as Sect.~\ref{sec:twonucleons} demonstrated, namely the deuteron
kinetic energy and $\Delta^{(0)}_d$ insertions depicted in
Fig.~\ref{fig:NLOthreebodyquartet}.
One might note that the diagrams are all finite thanks to the re-formulation
of the original Lagrangean (\ref{ksw}) into (\ref{dlag}) and hence can be
treated numerically~\cite{pbhg}. Each contribution is inserted only once.
Because of the absence of interactions which mix different partial waves, the
NLO contributions are given by
\begin{eqnarray}
  \label{insertion4dim}
  \ii t^{(l)}_{1\,\mathrm{q}}(k)&=& \int \frac{d^4q}{(2\pi)^4}\;
  (\ii t^{(l)}_{0\,\mathrm{q}}(k,q))^2\;
  \frac{\ii}{\frac{k^2}{2M}-q_0-\frac{q^2}{2M}+\ii\varepsilon}\;\times\\
  &&\phantom{\int \frac{d^4q}{(2\pi)^4}\;}\times\;
   \left[ \ii\triangle_d(\frac{
   k^2}{4M}-\frac{\gamma^2_d}{M}+q_0,\qv)\right]^2
 \;\ii\calI_d (\frac{k^2}{4M}-\frac{\gamma^2_d}{M}+q_0,\qv) \;\;,\nonumber
\end{eqnarray}
where the corrections to the deuteron are
\begin{equation}
  \label{insertion}
  \ii\calI_d(p_0,\pv)=-\ii\Delta^{(0)}_d-\ii(p_0-\frac{
   \pv^2}{4M})\;\;.
\end{equation}
The $q_0$ integration picks the nucleon pole resulting in a one dimensional
integral
\begin{eqnarray}\label{insertion2}
   \ii t^{(l)}_{1\,\mathrm{q}}(k)&=&\int\limits_0^\infty
   \frac{dq}{2\pi^2}\;q^2\;(\ii t^{(l)}_{0\,\mathrm{q}}(k,q))^2
   \left[ \ii\triangle_d(\frac{3
   k^2}{4M}-\frac{\gamma^2_d}{M}-\frac{q^2}{2M},\qv)\right]^2\;\times\\
   &&\phantom{\int\limits_0^\infty
   \frac{dq}{2\pi^2}}\;\times\;
   \ii\calI_d(\frac{3k^2}{4M}-\frac{\gamma^2_d}{M}-\frac{q^2}{2M},\qv)\nonumber
   \;\;,
\end{eqnarray}
which has to be performed numerically since $t^{(l)}_{0\,\mathrm{q}}(k,p)$ is
known only numerically.

Finally, the wave function renormalisation constant $Z$ at NLO is found from
\begin{eqnarray}
  \frac{1}{Z}&=&\frac{1}{Z_0+Z_1}\;\simeq\;\frac{1}{Z_0}-\frac{Z_1}{Z_0^2}
  \nonumber\\
  &=&\ii\frac{\de}{\de p_0}\;\frac{1}{\ii\triangle_d(p)+\ii\triangle_d(p)
  \ii\calI_d(p)\ii\triangle_d(p)}\bigg|_{p_0=\frac{\gamma^2_d}{M}\atop\pv=0}
  \;\simeq\;\frac{1}{Z_0}-\ii\;\frac{\partial}{\partial
    p_0}\; \ii\calI_d(p_0,\pv)\bigg|_{p_0=\frac{\gamma^2_d}{M}\atop\pv=0}
\end{eqnarray}
as
\begin{equation}\label{Z1}
  Z_1=Z_0^2\;\;.
\end{equation}
The NLO amplitude of the $l$th partial wave in the quartet channel is
therefore given by
\begin{eqnarray}
  T^{(l)}_{\mathrm{q}}(k)&=&Z\ t^{(l)}_{\mathrm{q}}(k,k)\nonumber\\ 
  &\simeq& (Z_0+Z_1)(t^{(l)}_{0\,\mathrm{q}}(k,k)+
  t^{(l)}_{1\,\mathrm{q}}(k,k))\simeq\nonumber\\ 
  &\simeq& T^{(l)}_{0\,\mathrm{q}}(k)+ Z_0 t^{(l)}_{1\,\mathrm{q}}(k,k)+Z_1
  t^{(l)}_{0\,\mathrm{q}}(k,k)=T^{(l)}_{0\,\mathrm{q}}(k)+
  T^{(l)}_{1\,\mathrm{q}}(k)\;\;.
\end{eqnarray}
The phase shift for each partial wave is extracted by expanding both sides of
the relation
\begin{equation}\label{kcotg}
  T^{(l)}_{\mathrm{q}}(k)\simeq T^{(l)}_{0\,\mathrm{q}}(k)+
  T^{(l)}_{1\,\mathrm{q}}(k)=\frac{3\pi}{M}\;
  \frac{1}{k\cot\delta^{(l)}(k) - ik}
\end{equation}
in $Q$ with $\delta^{(l)}=\delta^{(0)\,(l)}+\delta^{(1)\,(l)}+\dots$ and
keeping only linear terms,
\begin{eqnarray}
  \label{eq:deltas}
  \delta^{(0)\,(l)}&=& \frac{1}{2\ii}\ln\left(1+\frac{2\ii k M}{3\pi}
    T^{(l)}_{0\,\mathrm{q}}(k)\right)\;\;,\non\\
  \delta^{(1)\,(l)}&=&
  \frac{k M}{3\pi}\frac{T^{(l)}_{1\,\mathrm{q}}(k)}
  {\left(1+\frac{2\ii k M}{3\pi}T^{(l)}_{0\,\mathrm{q}}(k)\right)}\;\;.
\end{eqnarray}
At NNLO, there are two kinds of contributions. One comes from the same
deuteron kinetic energy and $\Delta^{(0)}$ terms appearing at NLO, but now
inserted twice. The other comes from inserting an operator describing the
tensor force which appears for example in the spin triplet channel of the
Lagrangean (\ref{ksw}) as
\begin{equation}
  \mathcal{L}_{NN,\,d}^{\mathrm{SD}}=\frac{1}{4}\; C_{2\,d}^{(\mathrm{SD})} 
  (N^T P^i N)^\dagger \left[N^T \left(P^j_d 
  (\stackrel{\scriptscriptstyle\rightarrow}{\de}-
      \stackrel{\scriptscriptstyle\leftarrow}{\de})^i
      (\stackrel{\scriptscriptstyle\rightarrow}{\de}-
      \stackrel{\scriptscriptstyle\leftarrow}{\de})^j
      -\frac{1}{3}P^i_d (\stackrel{\scriptscriptstyle\rightarrow}{\de}-
      \stackrel{\scriptscriptstyle\leftarrow}{\de})^2\right)N\right]+\;
      \mathrm{H.c.}
\end{equation}
and is generated by the term
\begin{equation}
  \mathcal{L}_{Nd}^{\mathrm{SD}}=y^{\mathrm{SD}}_d d^{i \dagger} \left[N
  \left((\stackrel{\scriptscriptstyle\rightarrow}{\de}-
      \stackrel{\scriptscriptstyle\leftarrow}{\de})^i
      (\stackrel{\scriptscriptstyle\rightarrow}{\de}-
      \stackrel{\scriptscriptstyle\leftarrow}{\de})^j 
      -\frac{1}{3}\delta^{ij} (\stackrel{\scriptscriptstyle\rightarrow}{\de}-
      \stackrel{\scriptscriptstyle\leftarrow}{\de})^2\right)P^j_d N\right]
    +\;\mathrm{H.c.}
\end{equation}
in the Lagrangean (\ref{dlag}) after the Gau\3ian integration over the
deuteron field is performed. This operator mixes spin and angular momentum and
hence produces a splitting and mixing oft amplitudes with the same spin $S$
and angular momentum $L$ but different values of the total angular momentum
$J$. This splitting is very important in determining spin observables.
However, although our calculation for the phase shifts themselves is at NNLO
and should be precise, the splittings start only at this order and should not
be very realistic. We postpone the analysis of these splittings to a future
publication where the spin observables will also be addressed and limit
ourselves here to the phase shifts averaged over different values of $J$. At
NNLO the tensor force appears only linearly, and a somewhat lengthy but
straightforward calculation shows that, after this average, its contribution
vanishes.

The following trick allows for a very simple calculation of the double
insertion of the kinetic and $\Delta^{(0)}$ operators. If they were included
to all orders, the only change from the LO calculation would be the use of the
NNLO deuteron propagator (\ref{n3loprops}) in (\ref{faddeevequationquartet})
instead of the LO one, (\ref{props}). This procedure re-sums some
contributions of order N3LO and higher, but not all of them, so it can be
justified only up to NNLO.  Therefore, a spin-averaged NNLO calculation is
simply performed by replacing the LO deuteron propagator (\ref{props}) by the
propagator (\ref{n3loprops}) which summarises all effective range effects, in
the integral equation (\ref{eqfort}).  As the resulting amplitude is unitary,
(\ref{kcotg}) is used directly to extract the phase shift. This procedure has
already been used in the calculation of the $\mathrm{S}$ wave quartet
scattering in~\cite{pbhg}.

\subsection{Doublet Channel}
\label{sec:doublet}
\begin{figure}[!htb]
  \begin{center}
    
    \vspace*{3ex}
    
    \setlength{\unitlength}{0.56pt}

%Insertions
    
    \feynbox{200\unitlength}{
            \begin{fmfgraph*}(200,64)
              \fmfleft{i2,i1} \fmfright{o2,o1}
              \fmf{double,width=thin,tension=3}{i1,v1}
              \fmf{double,width=thin,tension=1.5}{v1,v3,v2}
              \fmf{double,width=thin,tension=3}{v2,v5}
              \fmfv{decor.shape=cross,decor.size=6*thick}{v5}
              \fmf{double,width=thin,tension=3}{v5,v7}
              \fmf{double,width=thin,tension=1.5}{v7,v8,v9}
              \fmf{double,width=thin,tension=3}{v9,o1}
              \fmf{vanilla,width=thin}{i2,v4}
              \fmf{vanilla,width=thin,tension=0.5}{v4,v10}
              \fmf{vanilla,width=thin}{v10,o2} \fmffreeze \fmffreeze
              \fmf{ellipse,rubout=1,label=$\fs t_{0\,,\mathrm{d}}^{ij}$,
                label.dist=0.12w,label.side=right}{v3,v4}
              \fmf{ellipse,rubout=1,label=$\fs t_{0\,,\mathrm{d}}^{ij}$,
                label.dist=0.12w,label.side=left}{v8,v10}
              \fmf{vanilla,width=thin}{v3,v4} \fmf{vanilla,width=thin}{v8,v10}
              \end{fmfgraph*}}
            \hqm\hqm$+$\hqm\hqm \feynbox{200\unitlength}{
            \begin{fmfgraph*}(200,64)
              \fmfleft{i2,i1} \fmfright{o2,o1}
              \fmf{double,width=thin,tension=3}{i1,v1}
              \fmf{double,width=thin,tension=1.5}{v1,v3,v2}
              \fmf{double,width=thin,tension=3}{v2,v5}
              \fmfv{decor.shape=circle,decor.filled=empty, decor.size=3*thick,
                label=$\fs-\ii\Delta^{(0)}_d$,label.angle=90}{v5}
              \fmf{double,width=thin,tension=3}{v5,v7}
              \fmf{double,width=thin,tension=1.5}{v7,v8,v9}
              \fmf{double,width=thin,tension=3}{v9,o1}
              \fmf{vanilla,width=thin}{i2,v4}
              \fmf{vanilla,width=thin,tension=0.5}{v4,v10}
              \fmf{vanilla,width=thin}{v10,o2} \fmffreeze \fmffreeze
              \fmf{ellipse,rubout=1,label=$\fs t_{0\,,\mathrm{d}}^{ij}$,
                label.dist=0.12w,label.side=right}{v3,v4}
              \fmf{ellipse,rubout=1,label=$\fs t_{0\,,\mathrm{d}}^{ij}$,
                label.dist=0.12w,label.side=left}{v8,v10}
              \fmf{vanilla,width=thin}{v3,v4} \fmf{vanilla,width=thin}{v8,v10}
              \end{fmfgraph*}}
            \hqm\hqm$+$\hqm\hqm \feynbox{200\unitlength}{
            \begin{fmfgraph*}(200,64)
              \fmfleft{i2,i1} \fmfright{o2,o1}
              \fmf{double,width=thin,tension=3}{i1,v1}
              \fmf{double,width=thin,tension=1.5}{v1,v3}
              \fmf{dbl_dashes,width=thin,tension=1.5}{v3,v2}
              \fmf{dbl_dashes,width=thin,tension=3}{v2,v5}
              \fmfv{decor.shape=cross,decor.size=6*thick}{v5}
              \fmf{dbl_dashes,width=thin,tension=3}{v5,v7}
              \fmf{dbl_dashes,width=thin,tension=1.5}{v7,v8}
              \fmf{double,width=thin,tension=1.5}{v8,v9}
              \fmf{double,width=thin,tension=3}{v9,o1}
              \fmf{vanilla,width=thin}{i2,v4}
              \fmf{vanilla,width=thin,tension=0.5}{v4,v10}
              \fmf{vanilla,width=thin}{v10,o2} \fmffreeze \fmffreeze
              \fmf{ellipse,rubout=1,label=$\fs t_{0\,,\mathrm{d}}^{iA}$,
                label.dist=0.12w,label.side=right}{v3,v4}
              \fmf{ellipse,rubout=1,label=$\fs t_{0\,,\mathrm{d}}^{iA}$,
                label.dist=0.12w,label.side=left}{v8,v10}
              \fmf{vanilla,width=thin}{v3,v4} \fmf{vanilla,width=thin}{v8,v10}
              \end{fmfgraph*}}
            \hqm\hqm$+$\hqm\hqm \feynbox{200\unitlength}{
            \begin{fmfgraph*}(200,64)
              \fmfleft{i2,i1} \fmfright{o2,o1}
              \fmf{double,width=thin,tension=3}{i1,v1}
              \fmf{double,width=thin,tension=1.5}{v1,v3}
              \fmf{dbl_dashes,width=thin,tension=1.5}{v3,v2}
              \fmf{dbl_dashes,width=thin,tension=3}{v2,v5}
              \fmfv{decor.shape=circle,decor.filled=empty, decor.size=3*thick,
                label=$\fs-\ii\Delta^{(0)}_t$,label.angle=90}{v5}
              \fmf{dbl_dashes,width=thin,tension=3}{v5,v7}
              \fmf{dbl_dashes,width=thin,tension=1.5}{v7,v8}
              \fmf{double,width=thin,tension=1.5}{v8,v9}
              \fmf{double,width=thin,tension=3}{v9,o1}
              \fmf{vanilla,width=thin}{i2,v4}
              \fmf{vanilla,width=thin,tension=0.5}{v4,v10}
              \fmf{vanilla,width=thin}{v10,o2} \fmffreeze \fmffreeze
              \fmf{ellipse,rubout=1,label=$\fs t_{0\,,\mathrm{d}}^{iA}$,
                label.dist=0.12w,label.side=right}{v3,v4}
              \fmf{ellipse,rubout=1,label=$\fs t_{0\,,\mathrm{d}}^{iA}$,
                label.dist=0.12w,label.side=left}{v8,v10}
              \fmf{vanilla,width=thin}{v3,v4} \fmf{vanilla,width=thin}{v8,v10}
              \end{fmfgraph*}}
  \end{center}
    \caption{\label{fig:NLOthreebodydoublet} \sl The NLO contributions to $nd$
      scattering in the doublet channel.}
\end{figure}
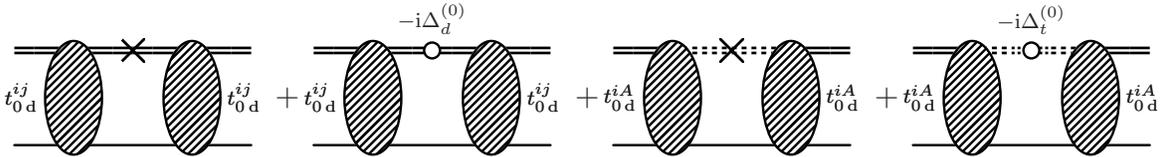

The calculation for the spin $1/2$ channels proceeds in an analogous way to
the one in the spin $3/2$ channels discussed above. The main difference is
that the spin zero di-baryon field $t$ also contributes in intermediate states
of $nd$ amplitudes. The analogue of (\ref{faddeevequationquartet}) is then a
system of two coupled integral equations for the $Nd\rightarrow Nd$ amplitude
$\ii t_{0\,\mathrm{d},\,Nd\to Nd}^{ij}(\vec{k},\vec{p},\epsilon)$ and for the
$Nd\rightarrow Nt$ amplitude $\ii t_{0\,\mathrm{d},\,Nd\to
  Nt}^{iA}(\vec{k},\vec{p},\epsilon)$, pictorially represented in
Fig.~\ref{fig:LOfaddeevdoublet}:
\begin{figure}[!htb]
  \begin{center}
    
    \vspace*{3ex}
    
    \setlength{\unitlength}{0.6pt}
    
    \feynbox{104\unitlength}{
            \begin{fmfgraph*}(104,64)
              \fmfleft{i2,i1} \fmfright{o2,o1}
              \fmf{double,width=thin,tension=3}{i1,v1}
              \fmf{double,width=thin,tension=1.5}{v1,v3,v2}
              \fmf{double,width=thin,tension=3}{v2,o1}
              \fmf{vanilla,width=thin}{i2,v4,o2} \fmffreeze
              \fmf{ellipse,rubout=1,label=$\fs t_{0\,,\mathrm{d}}^{ij}$,
                label.dist=0.25w,label.side=right}{v3,v4}
              \end{fmfgraph*}}
            \hq$=$\hq \feynbox{104\unitlength}{
            \begin{fmfgraph*}(104,64)
              \fmfleft{i2,i1} \fmfright{o2,o1}
              \fmf{double,width=thin,tension=4}{i1,v1,v2}
              \fmf{vanilla,width=thin}{v2,o1}
              \fmf{double,width=thin,tension=4}{v3,v4,o2}
              \fmf{vanilla,width=thin}{i2,v3} \fmffreeze
              \fmf{vanilla,width=thin}{v2,v3}
              \end{fmfgraph*}}
            \hq$+$\hq \feynbox{192\unitlength}{
            \begin{fmfgraph*}(192,64)
              \fmfleft{i2,i1} \fmfright{o2,o1}
              \fmf{double,width=thin,tension=3}{i1,v1}
              \fmf{double,width=thin,tension=1.5}{v1,v6,v5}
              \fmf{double,width=thin,tension=3}{v5,v2}
              \fmf{vanilla,width=thin}{v2,o1} \fmf{vanilla,width=thin}{i2,v7}
              \fmf{vanilla,width=thin,tension=0.666}{v7,v4}
              \fmf{double,width=thin,tension=4}{v4,v3,o2} \fmffreeze
              \fmf{vanilla,width=thin}{v4,v2} \fmf{ellipse,rubout=1,label=$\fs
                t_{0\,,\mathrm{d}}^{ij}$,
                label.dist=0.15w,label.side=right}{v6,v7}
              \end{fmfgraph*}}
            \hq$+$\hq \feynbox{192\unitlength}{
            \begin{fmfgraph*}(192,64)
              \fmfleft{i2,i1} \fmfright{o2,o1}
              \fmf{double,width=thin,tension=3}{i1,v1}
              \fmf{double,width=thin,tension=1.5}{v1,v6}
              \fmf{dbl_dashes,width=thin,tension=1.5}{v6,v5}
              \fmf{dbl_dashes,width=thin,tension=3}{v5,v2}
              \fmf{vanilla,width=thin}{v2,o1} \fmf{vanilla,width=thin}{i2,v7}
              \fmf{vanilla,width=thin,tension=0.666}{v7,v4}
              \fmf{double,width=thin,tension=4}{v4,v3,o2} \fmffreeze
              \fmf{vanilla,width=thin}{v4,v2} \fmf{ellipse,rubout=1,label=$\fs
                t_{0\,,\mathrm{d}}^{iA}$,
                label.dist=0.15w,label.side=right}{v6,v7}
              \end{fmfgraph*}}
            
            \vspace*{4ex}
            
            \feynbox{104\unitlength}{
            \begin{fmfgraph*}(104,64)
              \fmfleft{i2,i1} \fmfright{o2,o1}
              \fmf{double,width=thin,tension=3}{i1,v1}
              \fmf{double,width=thin,tension=1.5}{v1,v3}
              \fmf{dbl_dashes,width=thin,tension=1.5}{v2,v3}
              \fmf{dbl_dashes,width=thin,tension=3}{o1,v2}
              \fmf{vanilla,width=thin}{i2,v4,o2} \fmffreeze
              \fmf{ellipse,rubout=1,label=$\fs t_{0\,,\mathrm{d}}^{iA}$,
                label.dist=0.25w,label.side=right}{v3,v4}
              \end{fmfgraph*}}
            \hq$=$\hq \feynbox{104\unitlength}{
            \begin{fmfgraph*}(104,64)
              \fmfleft{i2,i1} \fmfright{o2,o1}
              \fmf{double,width=thin,tension=4}{i1,v1,v2}
              \fmf{vanilla,width=thin}{v2,o1}
              \fmf{dbl_dashes,width=thin,tension=4}{o2,v4,v3}
              \fmf{vanilla,width=thin}{i2,v3} \fmffreeze
              \fmf{vanilla,width=thin}{v2,v3}
              \end{fmfgraph*}}
            \hq$+$\hq \feynbox{192\unitlength}{
            \begin{fmfgraph*}(192,64)
              \fmfleft{i2,i1} \fmfright{o2,o1}
              \fmf{double,width=thin,tension=3}{i1,v1}
              \fmf{double,width=thin,tension=1.5}{v1,v6,v5}
              \fmf{double,width=thin,tension=3}{v5,v2}
              \fmf{vanilla,width=thin}{v2,o1} \fmf{vanilla,width=thin}{i2,v7}
              \fmf{vanilla,width=thin,tension=0.666}{v7,v4}
              \fmf{dbl_dashes,width=thin,tension=4}{o2,v3,v4} \fmffreeze
              \fmf{vanilla,width=thin}{v4,v2} \fmf{ellipse,rubout=1,label=$\fs
                t_{0\,,\mathrm{d}}^{ij}$,
                label.dist=0.15w,label.side=right}{v6,v7}
              \end{fmfgraph*}}
            \hq$+$\hq \feynbox{192\unitlength}{
            \begin{fmfgraph*}(192,64)
              \fmfleft{i2,i1} \fmfright{o2,o1}
              \fmf{double,width=thin,tension=3}{i1,v1}
              \fmf{double,width=thin,tension=1.5}{v1,v6}
              \fmf{dbl_dashes,width=thin,tension=1.5}{v6,v5}
              \fmf{dbl_dashes,width=thin,tension=3}{v5,v2}
              \fmf{vanilla,width=thin}{v2,o1} \fmf{vanilla,width=thin}{i2,v7}
              \fmf{vanilla,width=thin,tension=0.666}{v7,v4}
              \fmf{dbl_dashes,width=thin,tension=4}{o2,v3,v4} \fmffreeze
              \fmf{vanilla,width=thin}{v4,v2} \fmf{ellipse,rubout=1,label=$\fs
                t_{0\,,\mathrm{d}}^{iA}$,
                label.dist=0.15w,label.side=right}{v6,v7}
              \end{fmfgraph*}}
            
            \vspace*{1ex}
            
            \setlength{\unitlength}{1pt}
    
  \end{center}
    \caption{\label{fig:LOfaddeevdoublet} \sl The coupled set of Faddeev
      equations
      (\protect\ref{faddeevequationdoublet1}/\ref{faddeevequationdoublet2})
      which need to be solved for \protect$t_{0\,\mathrm{d}}$ at LO in the
      doublet channel. The double dashed line denotes the di-baryon field
      \protect$t^A$.}
\end{figure}
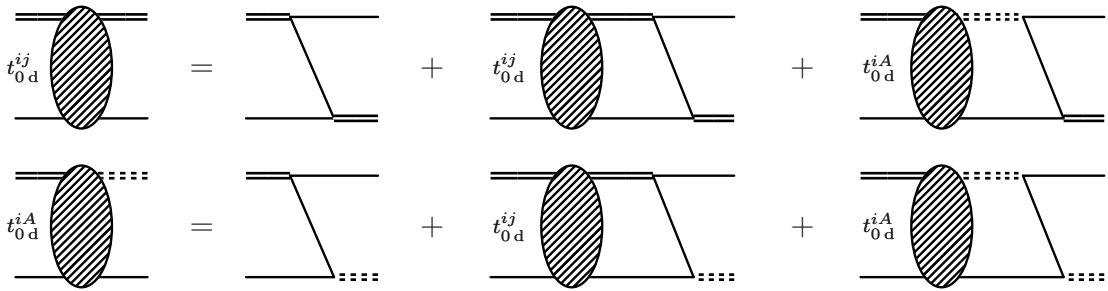
\begin{eqnarray}
  \label{faddeevequationdoublet1}
  \left(\ii t_{0\,\mathrm{d},\,Nd\to Nd}^{ij}\right)_{\alpha a}^{\beta b}
  (\kv,\pv,\epsilon) &=& \frac{y^2_d}{2}\;
  (\sigma^j\sigma^i)_\alpha^\beta \;\delta_a^b\;
  \frac{\ii}{-\frac{\kv^2}{4M}-\frac{\gamma^2_d}{M}+\epsilon
    -\frac{(\kv+\pv)^2}{2M}+ \ii\varepsilon}\;+\non\\
  &  &+\int
  \frac{d^4q}{(2\pi)^4}\ \Big[\;4 y^2_d \;
  (\sigma^j\sigma^k)_\gamma^\beta \;\delta_c^b\;
  \left(\ii t^{ik}_{0\,\mathrm{d},\,Nd\to Nd}\right)_{\alpha a}^{\gamma
    c}(\kv,\qv,\epsilon+q_0)\;\times\nonumber\\
 &  &\phantom{\int\frac{d^4q}{(2\pi)^4}\ \Big[\;4 y^2_d}\times
  \ii\triangle_d(\frac{\kv^2}{4}-\frac{\gamma^2_d}{M}+\epsilon+q_0,\qv)+
 \\
 &  &\phantom{\int
  \frac{d^4q}{(2\pi)^4} \ \Big[\;}+\;4 y_d y_t\;
  (\sigma^j)_\gamma^\beta \;(\tau^C)_c^b\; 
  \left(\ii t^{iC}_{0\,\mathrm{d},\,Nd\to Nt}\right)_{\alpha a}^{\gamma
    c}(\kv,\qv,\epsilon+q_0)\;\times\nonumber\\
 &  &\phantom{\int
  \frac{d^4q}{(2\pi)^4}\ \Big[\;4y_dy_t }\times
  \ii\triangle_t(\frac{\kv^2}{4}-\frac{\gamma^2_d}{M}+\epsilon+q_0,\qv)\Big]
  \;\times
  \nonumber\\
  & &\phantom{\int }% \frac{d^4q}{(2\pi)^4}\ }
  \times\;
  \frac{\ii}{\frac{\kv^2}{2M}-\epsilon-q_0-\frac{\qv^2}{2M}+ \ii\varepsilon}\
  \frac{\ii}{-\frac{\kv^2}{4M}-\frac{\gamma^2_d}{M}+2\epsilon
    +q_0-\frac{(\qv+\pv)^2}{2M}+\ii\varepsilon}\nonumber
\end{eqnarray}
\begin{eqnarray}
  \label{faddeevequationdoublet2}
  \left(\ii t^{iA}_{0\,\mathrm{d},\,Nd\to Nt}\right)_{\alpha a}^{\beta b}
  (\kv,\pv,\epsilon) &=& \frac{y_d y_t}{2}\;
  (\sigma^i)_\alpha^\beta \;(\tau^A)_a^b\;
  \frac{\ii}{-\frac{\kv^2}{4M}-\frac{\gamma^2_d}{M}+\epsilon
    -\frac{(\kv+\pv)^2}{2M}+ \ii\varepsilon}\;+\non\\
  &  &+\int
  \frac{d^4q}{(2\pi)^4}\  \Big[  \;4 y^2_t\;
  \delta_\gamma^\beta \;(\tau^A\tau^C)_c^b\; 
  \left(\ii t^{iC}_{0\,\mathrm{d},\,Nd\to Nt}\right)_{\alpha a}^{\gamma
    c}(\kv,\qv,\epsilon+q_0)\;\times\nonumber\\
 &  &\phantom{\int \frac{d^4q}{(2\pi)^4} \  \Big[ \;4 y^2_t}\times
  \ii\triangle_t(\frac{\kv^2}{4}-\frac{\gamma^2_d}{M}+\epsilon+q_0,\qv)\;+
  \\
 &  &\phantom{\int
  \frac{d^4q}{(2\pi)^4}\  \Big[ \;}+\;4 y_d y_t\;
  (\sigma^k)_\gamma^\beta \;(\tau^A)_c^b\; 
  \left(\ii t^{ik}_{0\,\mathrm{d},\,Nd\to Nd}\right)_{\alpha a}^{\gamma
    c}(\kv,\qv,\epsilon+q_0)\;\times\nonumber\\
 &  &\phantom{\int\frac{d^4q}{(2\pi)^4} \  \Big[+\;4 y_d}
 \times
  \ii\triangle_d(\frac{\kv^2}{4}-\frac{\gamma^2_d}{M}+\epsilon+q_0,\qv)\Big]
  \;\times\nonumber\\
  & &  \phantom{\int}%\frac{d^4q}{(2\pi)^4}\ }
  \times\;
 \frac{\ii}{\frac{\kv^2}{2M}-\epsilon-q_0-\frac{\qv^2}{2M}+ \ii\varepsilon}\
  \frac{\ii}{-\frac{\kv^2}{4M}-\frac{\gamma^2_d}{M}+2\epsilon
    +q_0-\frac{(\qv+\pv)^2}{2M}+\ii\varepsilon}\nonumber
\end{eqnarray}
After projecting on the spin $1/2$ channel using
\begin{eqnarray}
  \left(t_{0\,\mathrm{d},\,Nd\to Nd}\right)_{\alpha a}^{\beta b}&=&
  \frac{1}{3} (\sigma^i)_\alpha^{\alpha'}
  \left(\ii t^{ij}_{0\,\mathrm{d},\,Nd\to Nd}\right)_{\alpha' a}^{\beta' b}
  (\sigma^j)_{\beta'}^{\beta}\;\;,\non\\
  \left(t_{0\,\mathrm{d},\,Nd\to Nt}\right)_{\alpha a}^{\beta b}&=&
  \frac{1}{3} (\sigma^i)_\alpha^{\alpha'} \left(\ii t^{iA}_{0\,\mathrm{d},
      \,Nd\to Nt}\right)_{\alpha' a}^{\beta b'} (\tau^A)_{b'}^{b}\;\;,
\end{eqnarray}
and following the same steps as in the quartet case one finds
\begin{eqnarray}
  \label{eqfort2}
  t^{(l)}_{0\,\mathrm{d},\,Nd\to Nd}(k,p)&=&\;\frac{4y^2_d M}{p k}\;
  Q_l\left({{p^2+k^2-ME
        - \ii\varepsilon} \over {pk}}\right)+ \non\\
  &&+\;\frac{1}{\pi}\int\limits_0^\infty dq\; q^2\;
  t^{(l)}_{0\,\mathrm{d},\,Nd\to Nd}(k,q)\;
  \frac{1}{\sqrt{\frac{3 q^2}{4}-ME-
     \ii\varepsilon}-\gamma_d}\;\times\nonumber\\
 &&\phantom{+\;\frac{1}{\pi}\int\limits_0^\infty dq\; q^2\;}\times\;
 \frac{1}{qp}\;
 Q_l\left({{p^2 + q^2-ME  - \ii\varepsilon} \over {pq}}\right)- \\
 &&-\;\frac{3}{\pi}\;\frac{y_d}{y_t}\int\limits_0^\infty dq\; q^2\;
 t^{(l)}_{0\,\mathrm{d},\,Nd\to Nt}(k,q)\;
 \frac{1}{\sqrt{\frac{3 q^2}{4}-ME- \ii\varepsilon}-\gamma_t}\;
 \times\nonumber\\
 &&\phantom{-\;\frac{3}{\pi}\;\frac{y_d}{y_t}\int\limits_0^\infty dq\; q^2\;}
 \times\;\frac{1}{qp}\;
 Q_l\left({{p^2 + q^2-ME  - \ii\varepsilon} \over {pq}}\right)\;\;,\nonumber
\end{eqnarray}
\begin{eqnarray}
  \label{eqfort3}
  t^{(l)}_{0\,\mathrm{d},\,Nd\to Nt}(k,p)&=&-\;\frac{12y_d y_t M}{p k}\;
  Q_l\left({{p^2+k^2-ME
        - \ii\varepsilon} \over {pk}}\right)+ \non\\
 &&+\;\frac{1}{\pi}\int\limits_0^\infty dq\; q^2\;
 t^{(l)}_{0\,\mathrm{d},\,Nd\to Nt}(k,q)\;
 \frac{1}{\sqrt{\frac{3 q^2}{4}-ME- \ii\varepsilon}-\gamma_t}\;
 \times\nonumber\\
 &&\phantom{+\;\frac{1}{\pi}\int\limits_0^\infty dq\; q^2\;}
 \times\;\frac{1}{qp}\;
 Q_l\left({{p^2 + q^2-ME  - \ii\varepsilon} \over
     {pq}}\right)- \\
 &&-\;\frac{3}{\pi}\;\frac{y_t}{y_d}\int\limits_0^\infty dq\; q^2\;
 t^{(l)}_{0\,\mathrm{d},\,Nd\to Nd}(k,q)\;
 \frac{1}{\sqrt{\frac{3 q^2}{4}-ME- \ii\varepsilon}-\gamma_d}
 \;\times\nonumber\\
 &&\phantom{-\;\frac{3}{\pi}\;\frac{y_t}{y_d}\int\limits_0^\infty dq\; q^2\;}
 \times\;\frac{1}{qp}\;
 Q_l\left({{p^2 + q^2-ME  - \ii\varepsilon} \over
     {pq}}\right)\;\;.\nonumber
\end{eqnarray}
The modifications for the NLO and NNLO calculation are straightforward.
Fig.~\ref{fig:NLOthreebodydoublet} shows the graphs giving the NLO
corrections. At NNLO, the propagators $\ii\triangle_{d/t}(p)$ in the above
integral equations are replaced by $\ii\triangle_{d/t}(p;\mathrm{N3LO})$ of
(\ref{n3loprops}).

%%%%%%%%%%%%%%%%%%%%%%%%%%%%%%%%%%%%%%%%%%%%%%%%%%%%%%%%%%%%%%%%%%%%%%%%%%%%%%%
\section{Discussion and Conclusions}
\label{sec:conclusion}
\setcounter{equation}{0}

\begin{figure}[!ph]
  \begin{center}
    \absatz \centerline{\includegraphics*[height=0.45\linewidth,
      angle=90]{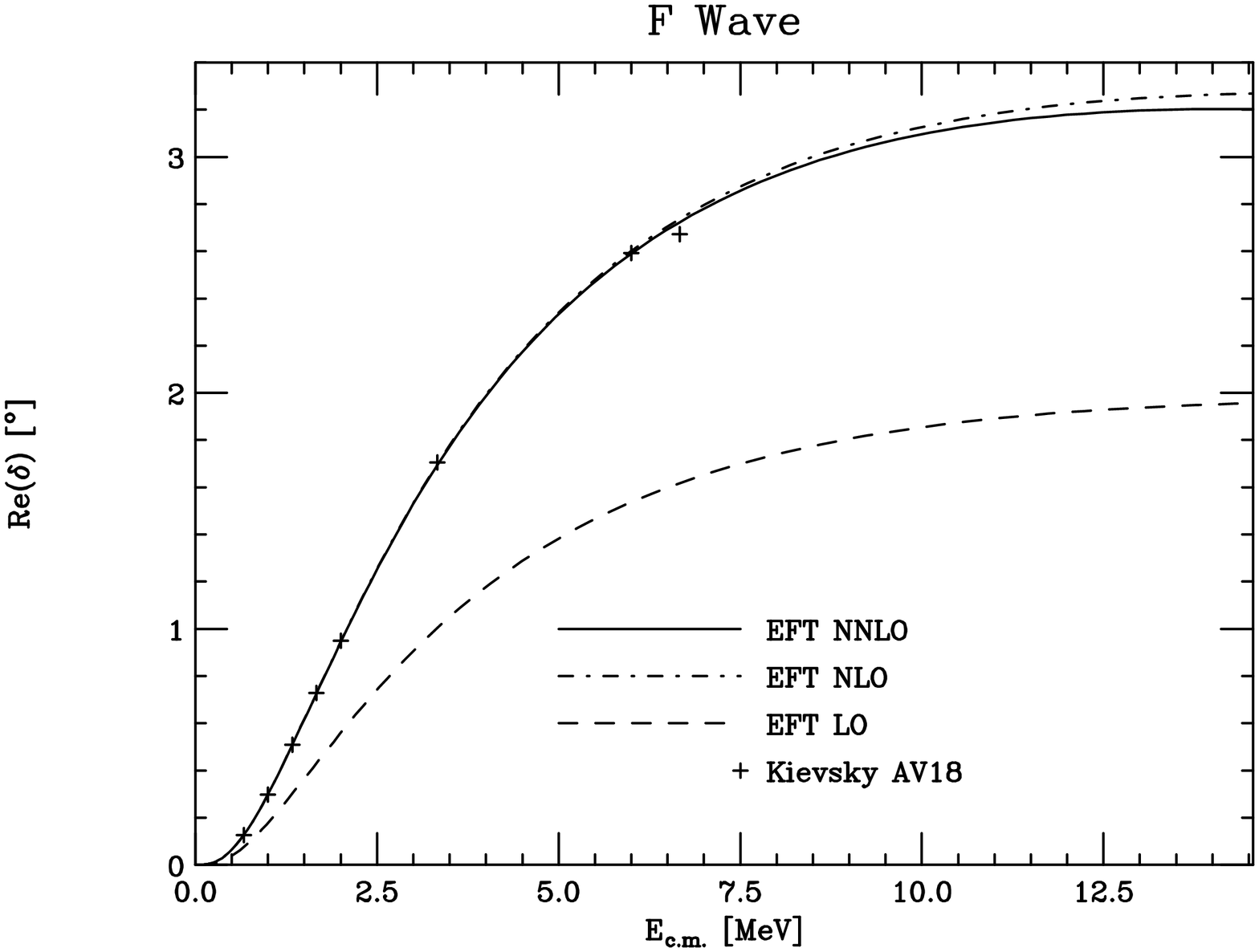} \hfill
      \includegraphics*[height=0.45\linewidth,angle=90]{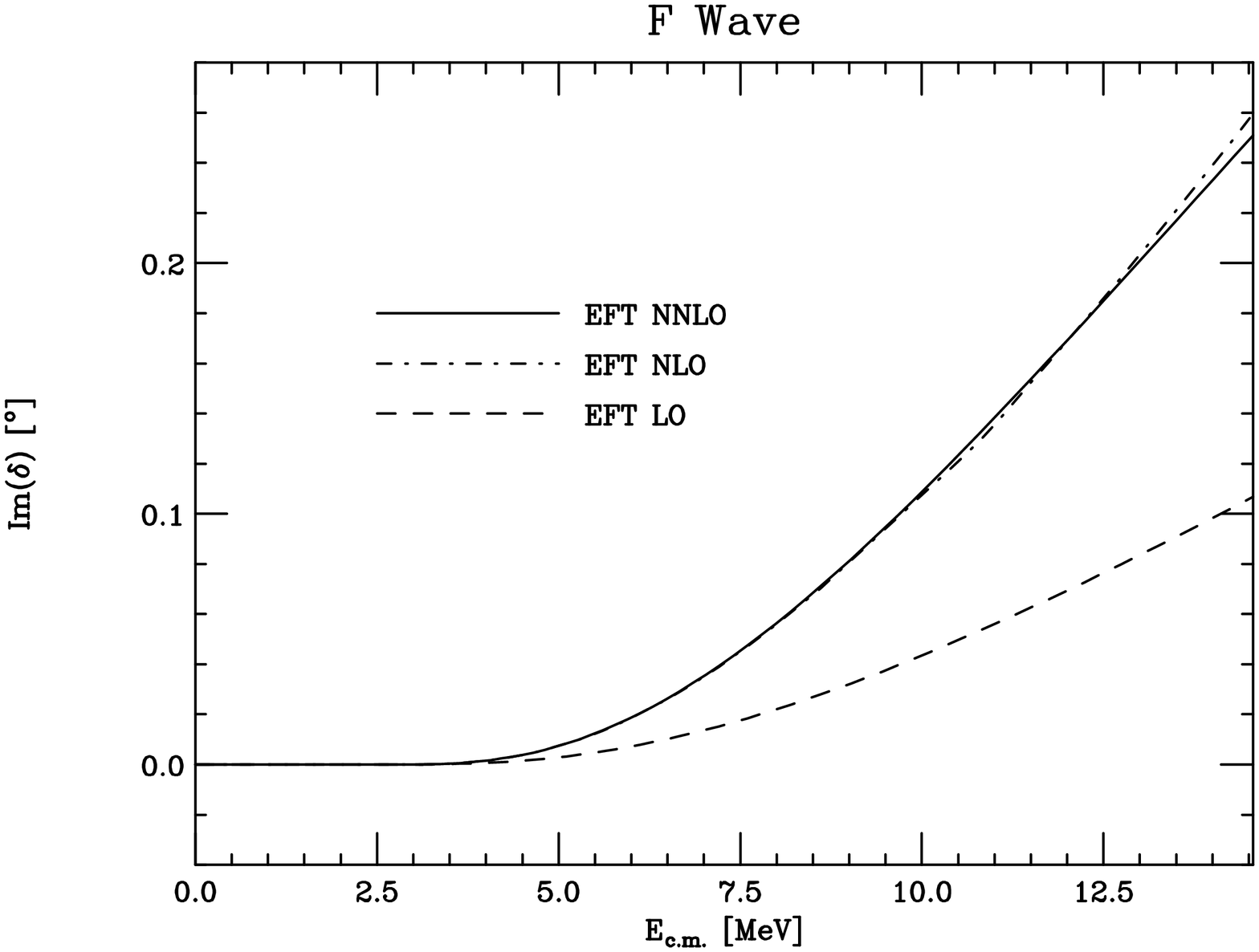} }
  
    \absatz \centerline{\includegraphics*[height=0.45\linewidth,
      angle=90]{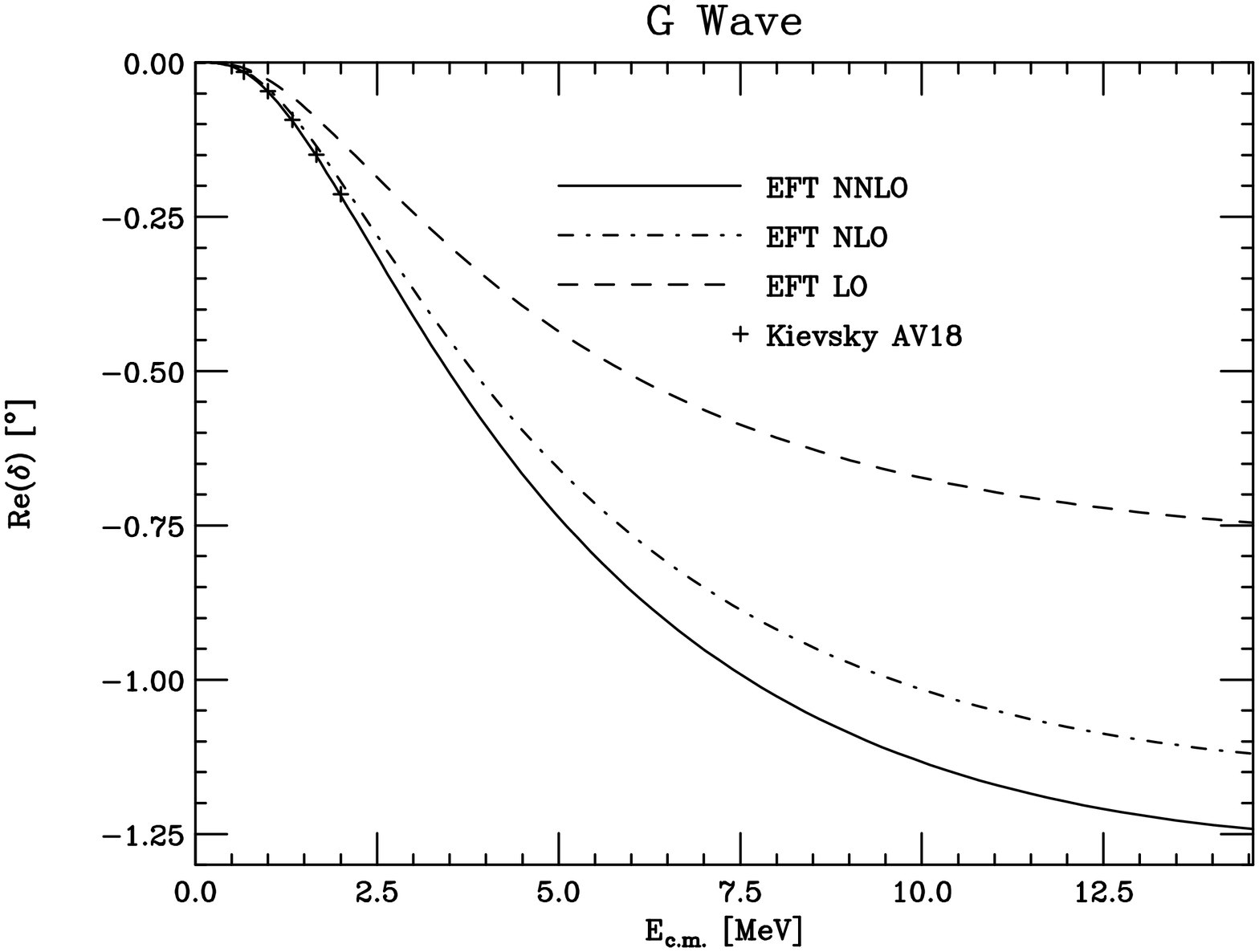} \hfill
      \includegraphics*[height=0.45\linewidth,angle=90]{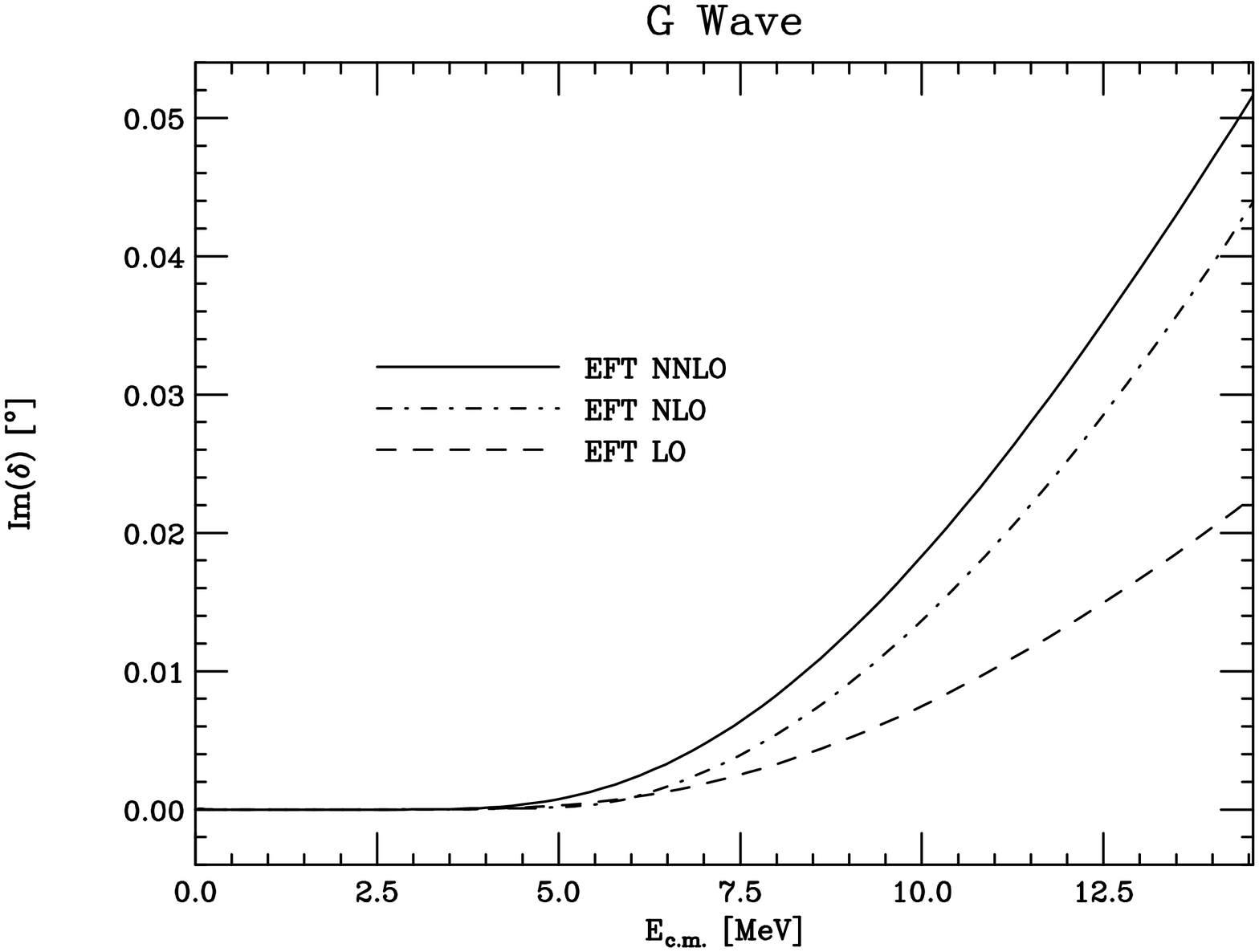} }
  
    \caption{\sl Real and imaginary parts of the first five partial waves in
      the quartet channel of \protect$nd$ scattering versus the centre-of-mass
      energy. The dashed line is the LO, the dot-dashed the NLO and the solid
      line the NNLO result. The calculations of Kievsky et al.~(crosses) are
      from Refs.~\protect\cite{Kievsky} below breakup (\protect$E_\mathrm{cm}
      =B$) and \protect\cite{Kievskyprivcomm} above breakup. The phase shift
      analyses of $pd$ data by Huttel et al.~\cite{Hutteletal} and Schmelzbach
      et al.~\cite{Schmelzbachetal} are presented as open squares and
      diamonds, respectively.}
    \label{fig:quartet}
  \end{center}
\end{figure}

\begin{figure}[!htp]
  \begin{center}
    \absatz \centerline{\includegraphics*[height=0.45\linewidth,
      angle=90]{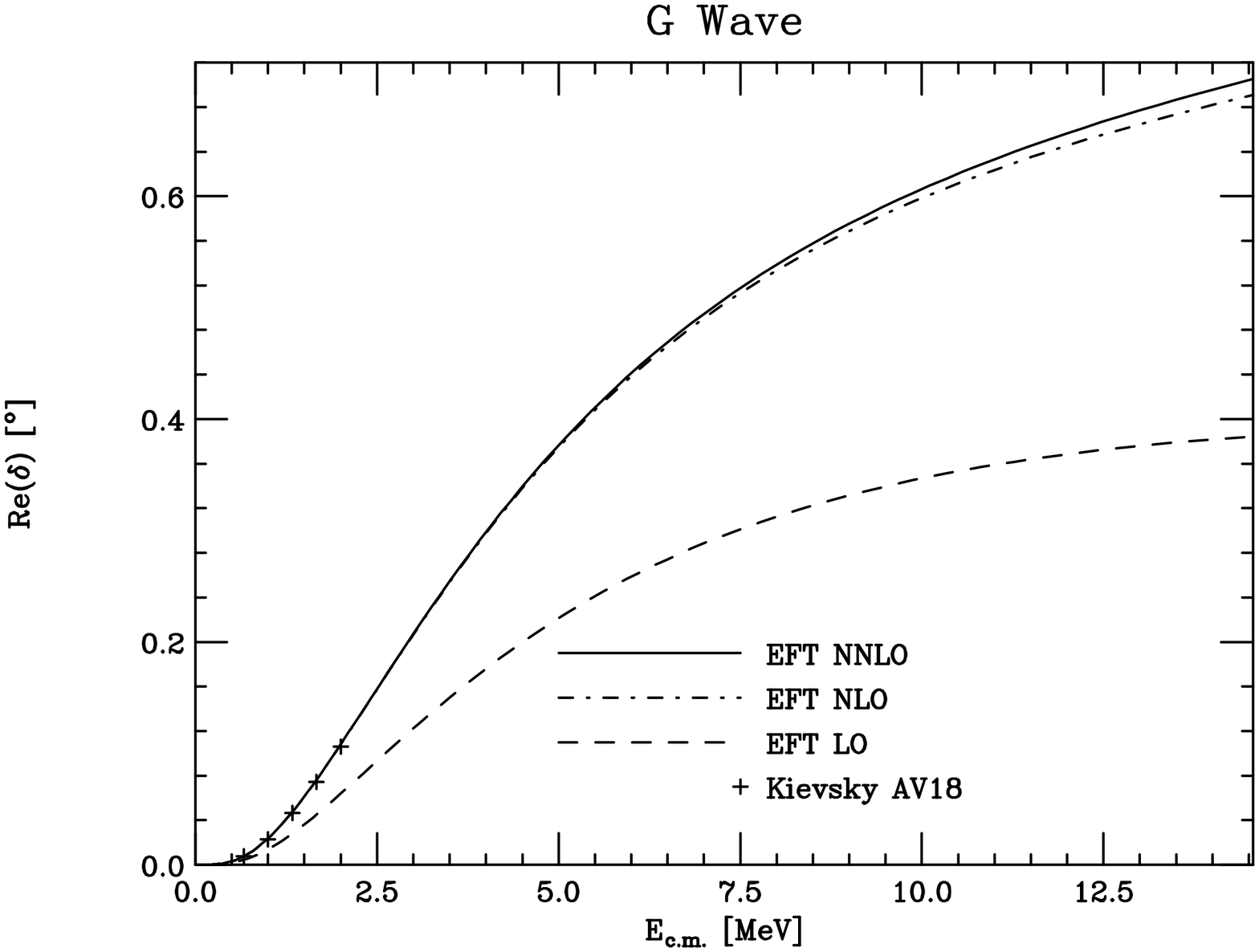} \hfill
      \includegraphics*[height=0.45\linewidth,angle=90]{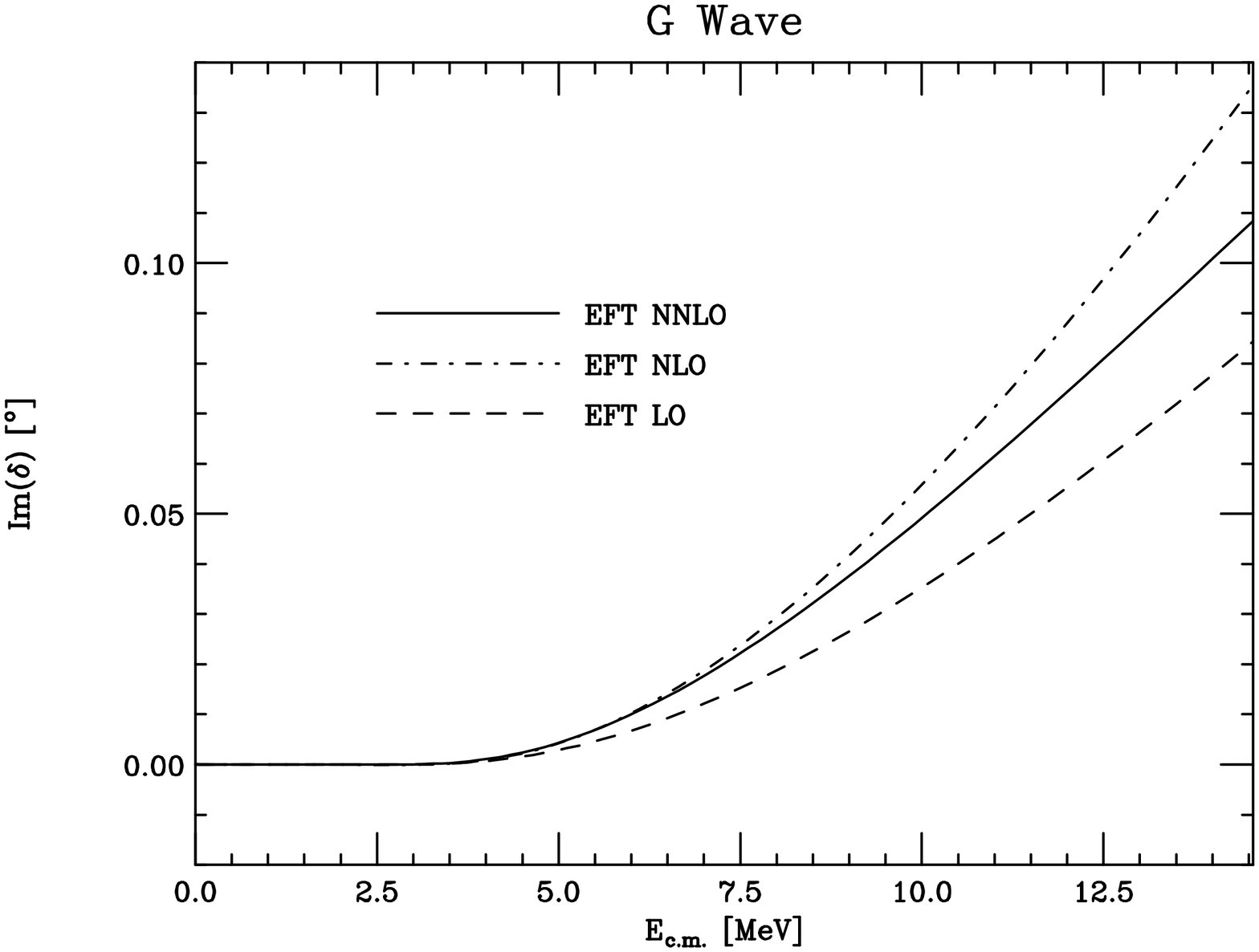} }
    \caption{\sl Real and imaginary parts of the first four higher partial
      waves in the doublet channel of \protect$nd$ scattering versus the
      centre-of-mass energy. Notation as in Fig.~\protect\ref{fig:quartet}.}
    \label{fig:doublet}
  \end{center}
\end{figure}

We used $\hbar c=197.327\;\MeV\,\fm$, a nucleon mass of $M=938.918\;\MeV$, for
the $NN$ triplet channel a deuteron binding energy (momentum) of
$B=2.225\;\MeV$ ($\gamma_d=45.7066\;\MeV$), a residue of $Z_d=1.690(3)$, and
for the $NN$ singlet channel an ${}^1\mathrm{S}_0$ scattering length of
$a_t=-23.714\;\fm$ and effective range $r_{0t}=2.73\;\fm$.  In
Fig.~\ref{fig:quartet}, we present the results of our computations for the
real and imaginary parts of the quartet partial waves $l=0$ to $4$ in the
centre-of-mass frame. Fig.~\ref{fig:doublet} shows the real and imaginary
parts of the doublet partial waves $l=1$ to $4$. The
${}^4\mathrm{S}_\frac{3}{2}$ partial wave was already computed in a previous
publication~\cite{pbhg} with a choice of parameters and method to extract the
phase shift which differs from the one used here formally only at higher order
in the power counting, and indeed the difference is marginal.  Comparison of
the LO with the NLO and NNLO result demonstrates convergence of the effective
field theory, with the expansion parameter found to be about $\frac{1}{3}$. We
therefore claim that the NNLO calculation has an error bar of less than $4$
percent. It is interesting to note that the NLO correction is for high enough
energies sometimes sizeable, while the NNLO correction is in general very
small. In contradistinction to the $NN$ case, the hierarchy the phase shifts
exhibit at large momenta with increasing $l$ is not due to suppression of the
higher partial waves in powers of $Q$, as the diagrams scale for all partial
waves in the same way. Rather, it is easy to see that especially for the
quartet channel, already the behaviour of the LO integral equations
(\ref{eqfort}) and (\ref{eqfort2}/\ref{eqfort3}) as $p=k\gg\gamma_{d/t}$ makes
the $l$th partial wave asymptotically go like the $l$th Legendre polynomial of
the second kind $Q_l(\frac{5}{4})$, and
$Q_l(\frac{5}{4})/Q_{l-1}(\frac{5}{4})\approx 0.33-0.45$, which is by accident
close to the expansion parameter $Q$.

Below the deuteron breakup point, our calculation is in good agreement with
the variational calculation of Kievsky et al.~\cite{Kievsky} which is based on
the AV18 potential. We also compare to partial wave analyses of $pd$
scattering experiments by Huttel et al.~\cite{Hutteletal} and Schmelzbach et
al.~\cite{Schmelzbachetal} as Coulomb and chiral symmetry breaking effects can
be neglected at high momenta. Above $E_\mathrm{cm}\sim 15\;\MeV$, i.e.~a
cm-momentum of about $140\;\MeV$, one expects the pion-less theory to diverge
from experiment because not later than then should the pion manifest itself as
an explicit degree of freedom of the low energy theory.

Within the range of validity of this pion-less theory we seem to have good
convergence and our results agree with potential model calculations within the
theoretical uncertainty of $(1/3)^3 \approx 4 \%$. That makes us optimistic
about carrying out higher order calculations of problematic spin observables
where our approach will differ from potential model calculations due to the
inclusion of three-body forces. For example, the nucleon-deuteron vector
analysing power $A_y$ in elastic $Nd$ scattering at energies below
$E_\mathrm{lab}\approx 30\;\MeV$ does not have a satisfactory description in
any realistic potential model of $NN$ scattering, see
\cite{HueberFriar,KievskyAy} for details. It remains to be seen whether the EFT
approach can be of help there.

We close with a note comparing the EFT approach at this order to realistic
potential model calculations. Since we use as input parameters to NNLO only
the scattering lengths $1/\gamma_d,\,a_t$ and effective ranges
$\rho_d,\,r_{0t}$ of the two body system in the $\mathrm{S}$ wave channels, we
expect any realistic potential model which reproduces these numbers to agree
with our results within the error bar of an NNLO calculation, $\sim 4\,\%$.
The absence of counterterms at this order (e.g.~of three body forces)
guarantees that an effective theory of a realistic potential model will be
identical to the EFT discussed here. Hence, a potential model can already
serve for the purpose of comparing to our results at energies considerably
higher than the deuteron breakup point. However, the EFT approach is
systematic and rigorous, allowing for error estimates. The small number and
simplicity of the interactions in the Lagrangean at NNLO makes the
computations considerably simpler: The LO integral equation is only
one-dimensional and can be obtained in very short computing time, and at NLO,
only (partially analytical, partially simple numerical) one dimensional
integrations are necessary, in contradistinction to many dimensional integral
equations in traditional approaches. Calculations of phase shifts in any
approach, be it potential models or EFT with pions as explicit degrees of
freedom, are hence bound to reproduce our results within error-bars in the
range of validity of the pion-less theory, allowing for valuable cross-checks.
Finally, rigorously determined three body forces will enter at higher orders
of the EFT approach and lift the similarity to potential model calculations.

\begin{figure}[!ph]
  \begin{center}
    \absatz \centerline{\includegraphics*[height=0.45\linewidth,
      angle=90]{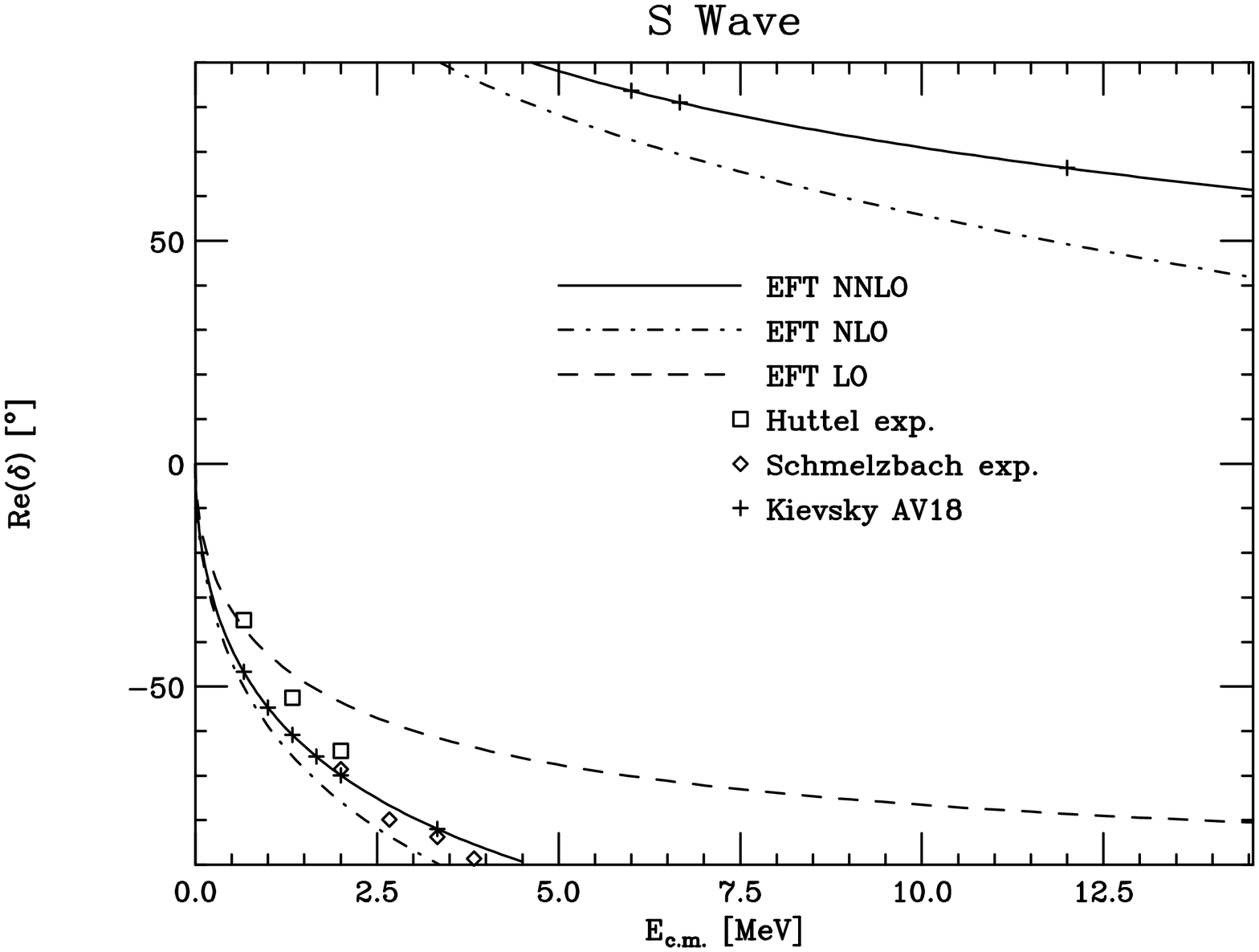} \hfill
      \includegraphics*[height=0.45\linewidth,angle=90]{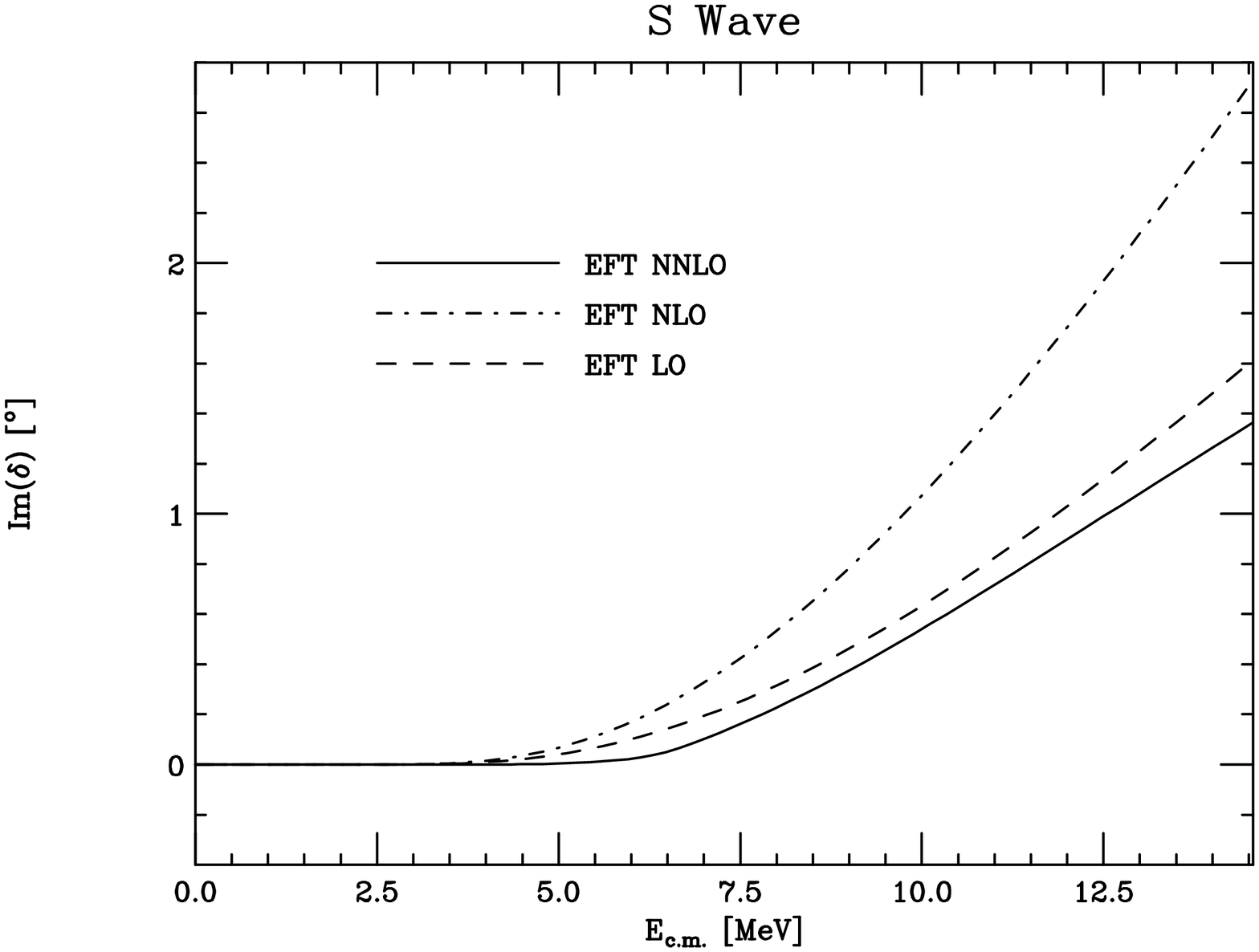} }
    
    \absatz \centerline{\includegraphics*[height=0.45\linewidth,
      angle=90]{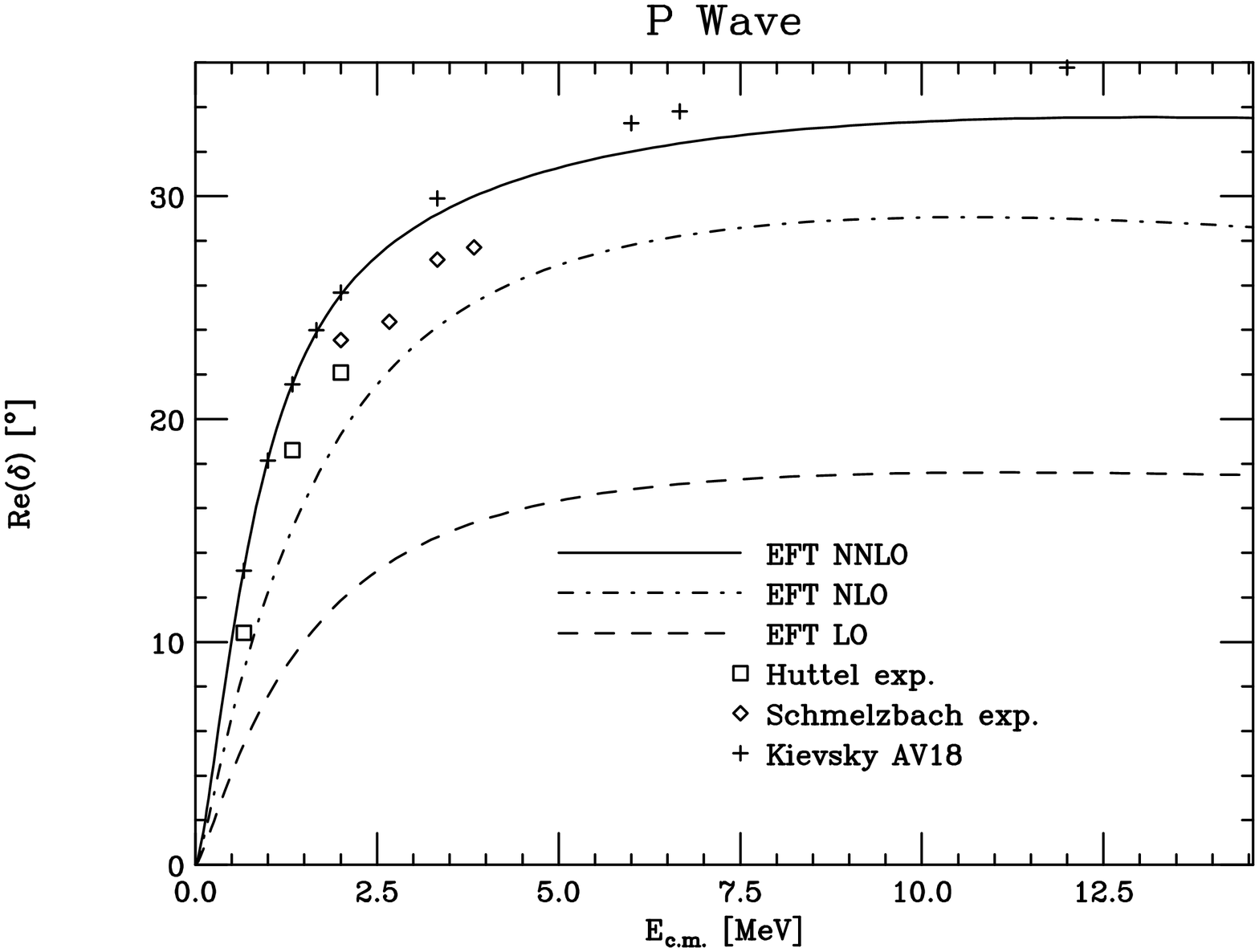} \hfill
      \includegraphics*[height=0.45\linewidth,angle=90]{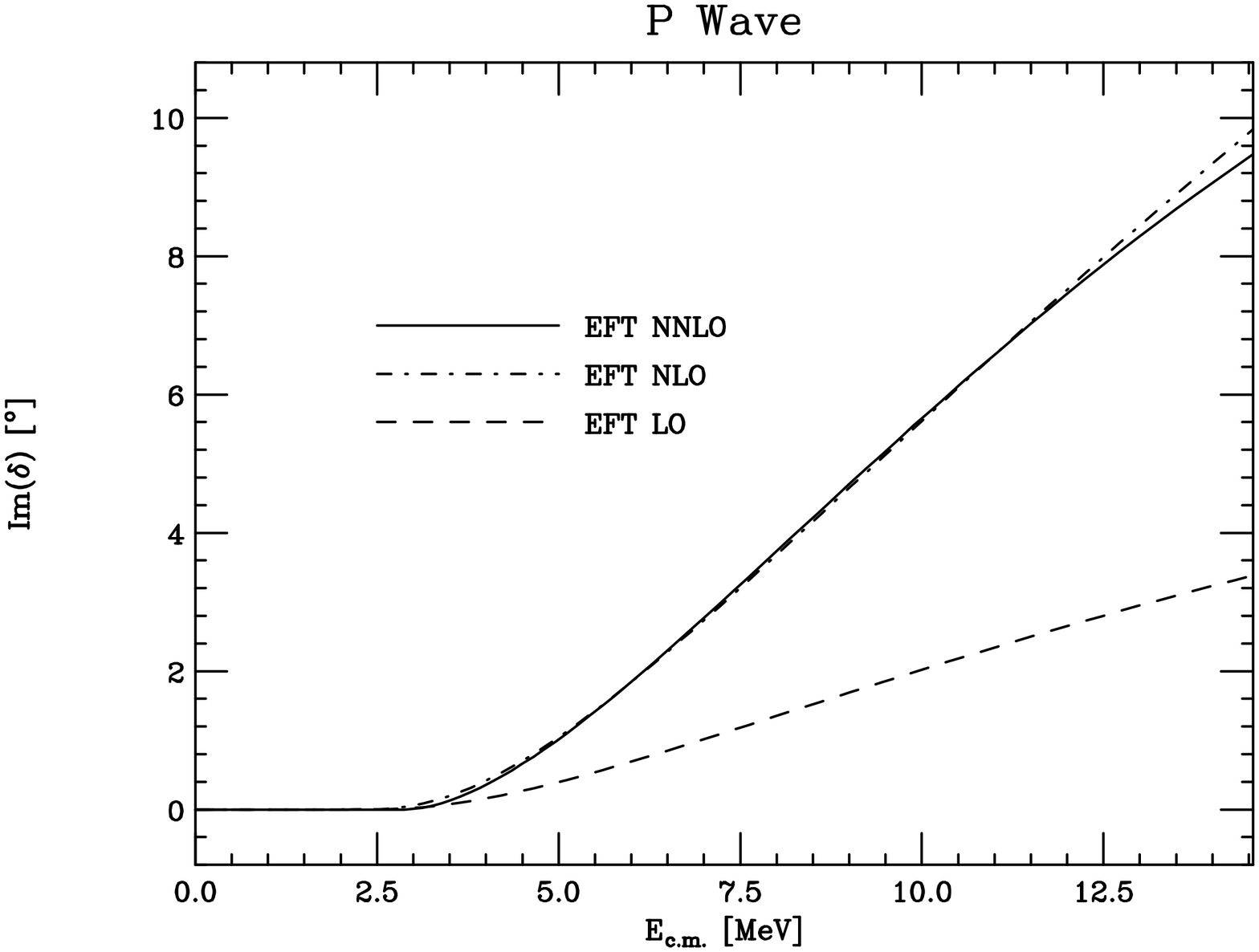} }
    
    \absatz \centerline{\includegraphics*[height=0.45\linewidth,
      angle=90]{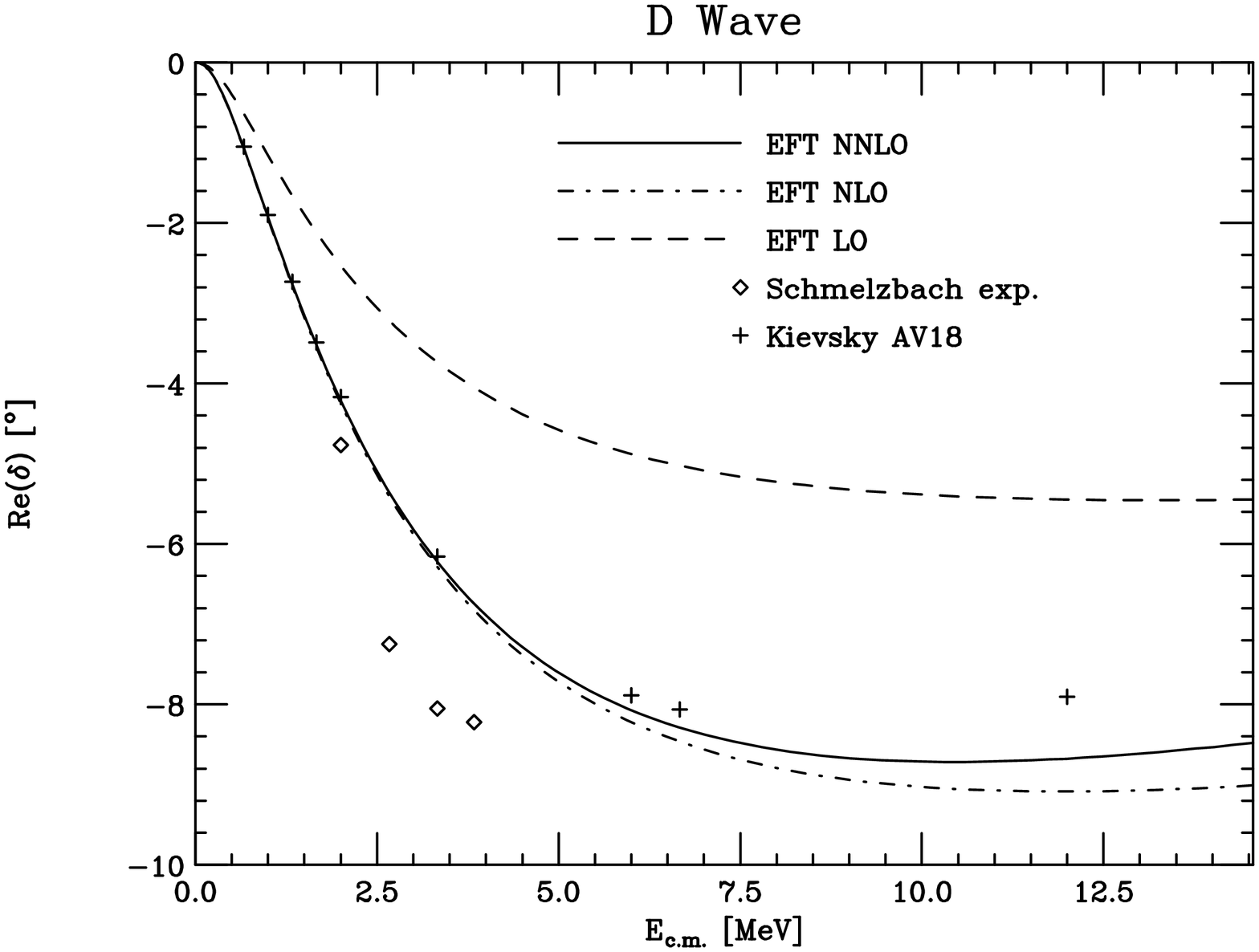} \hfill
      \includegraphics*[height=0.45\linewidth,angle=90]{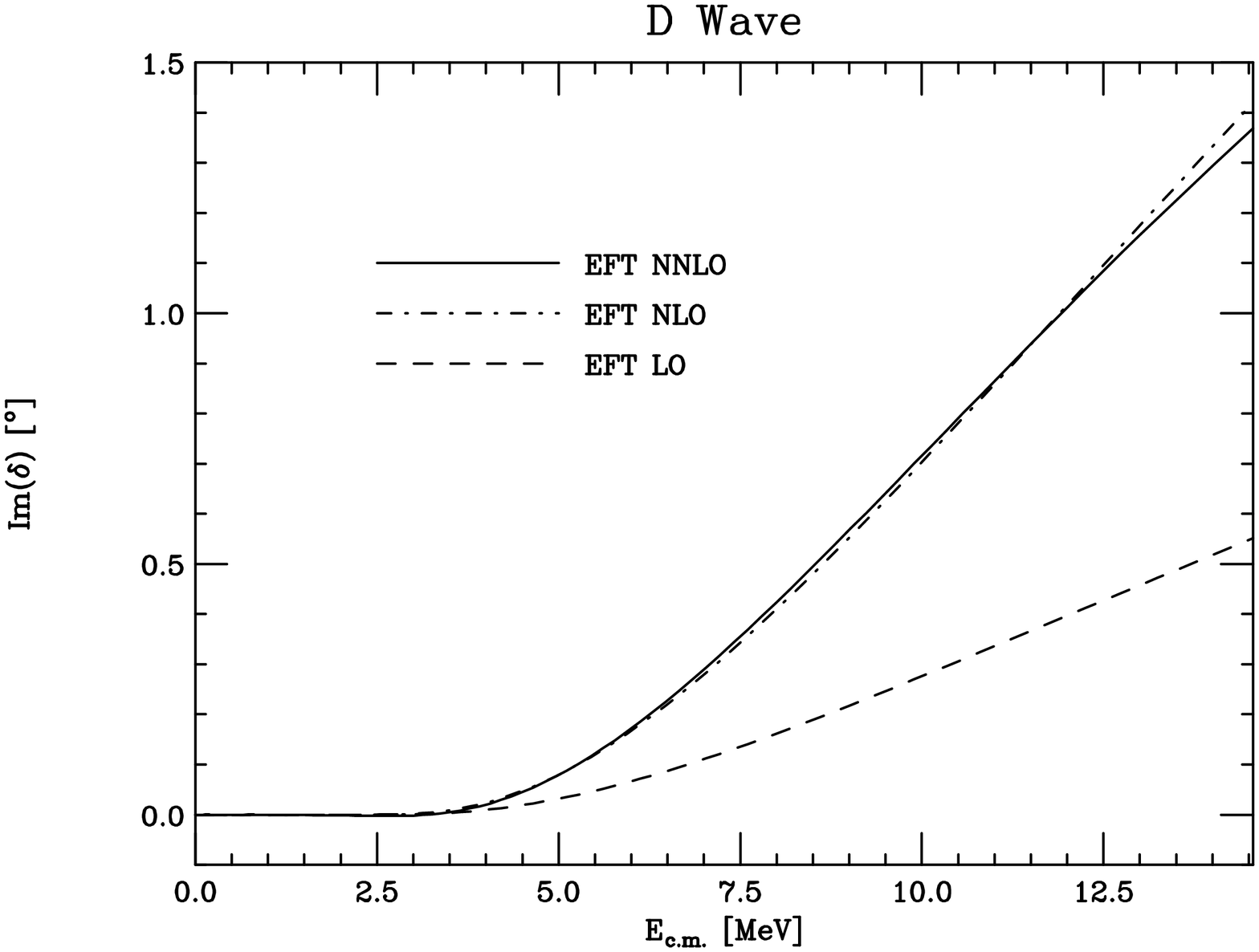} }
  \end{center}
\end{figure}
\begin{figure}[!ph]
  \begin{center}
    \absatz \centerline{\includegraphics*[height=0.45\linewidth,
      angle=90]{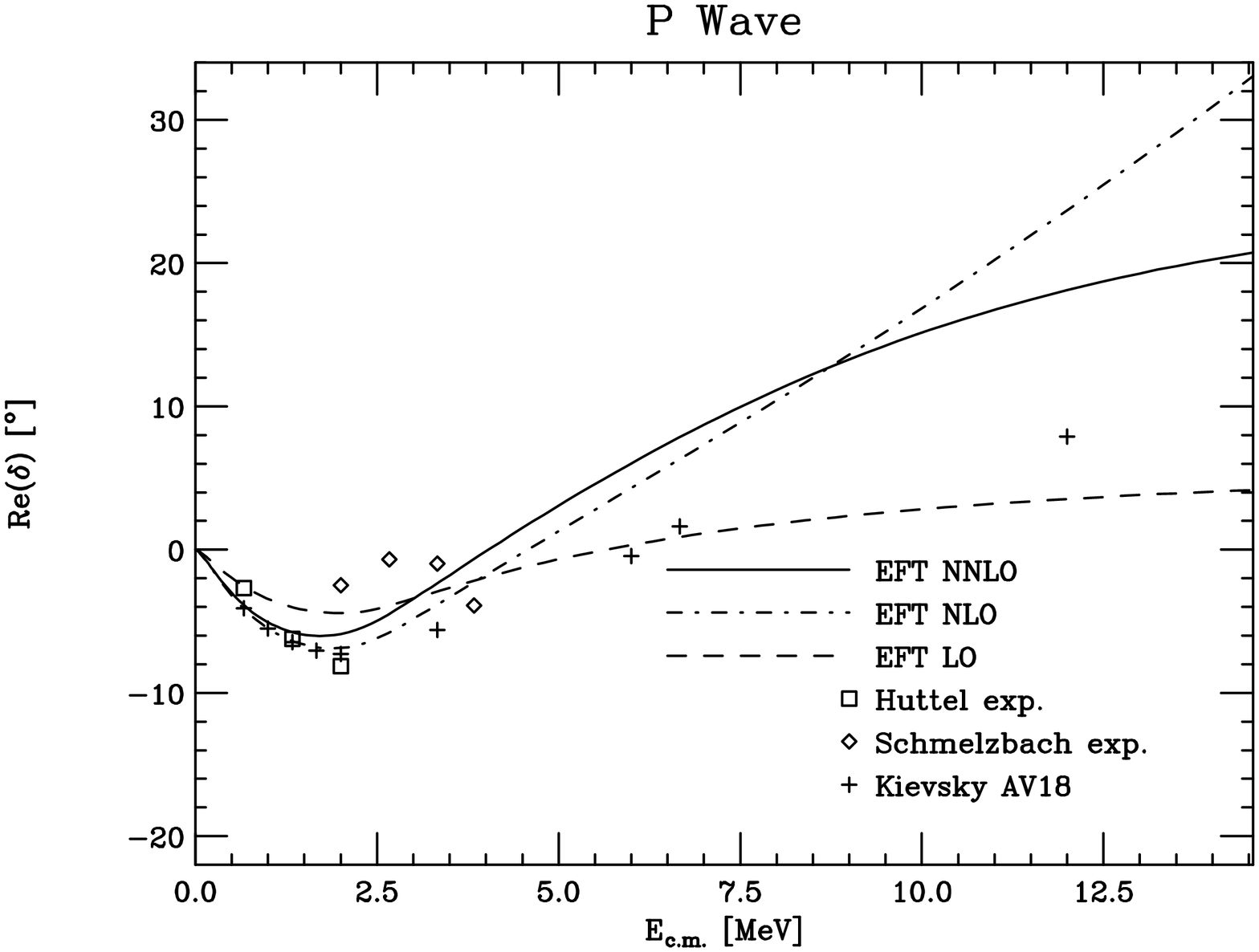} \hfill
      \includegraphics*[height=0.45\linewidth,angle=90]{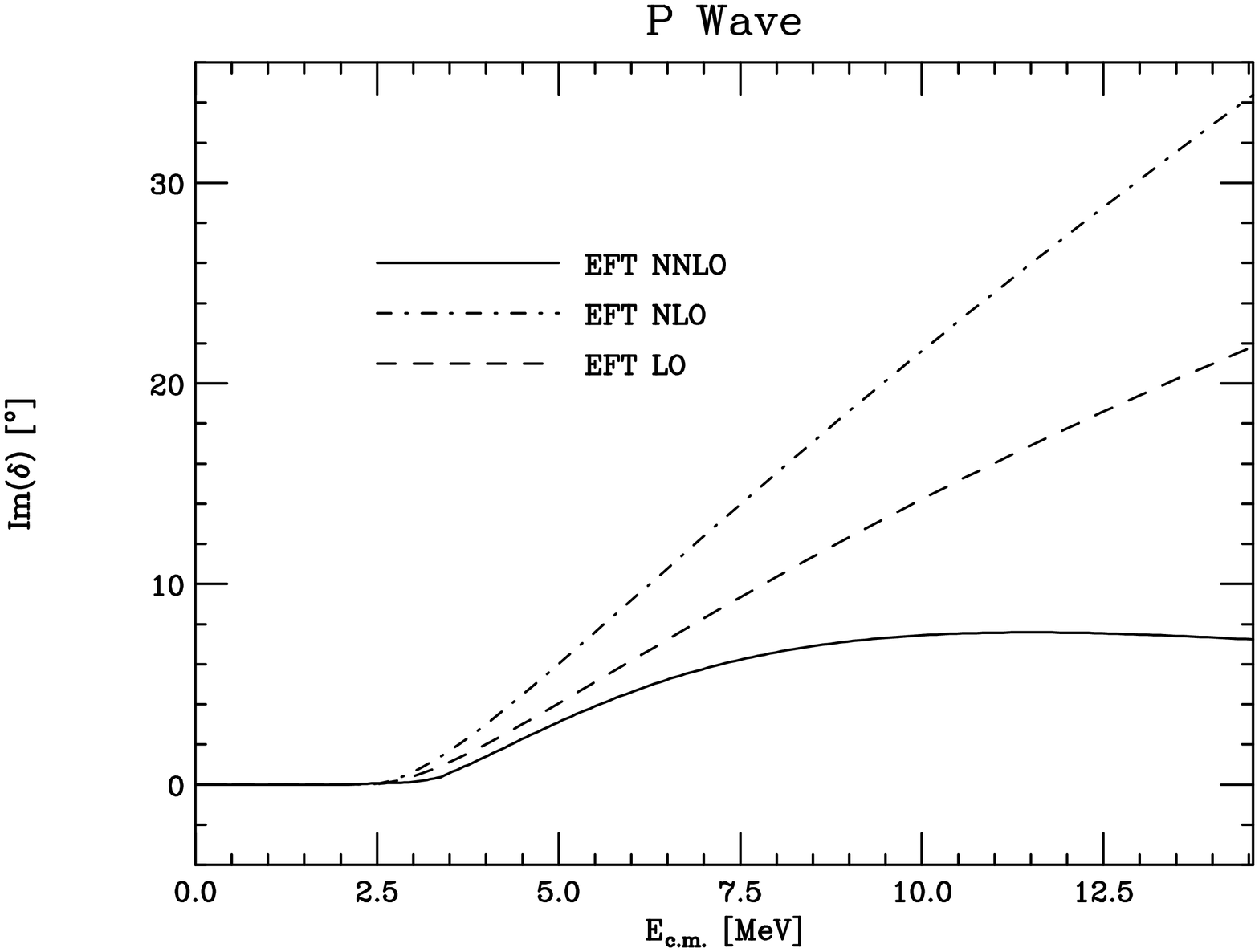} }
    
    \absatz \centerline{\includegraphics*[height=0.45\linewidth,
      angle=90]{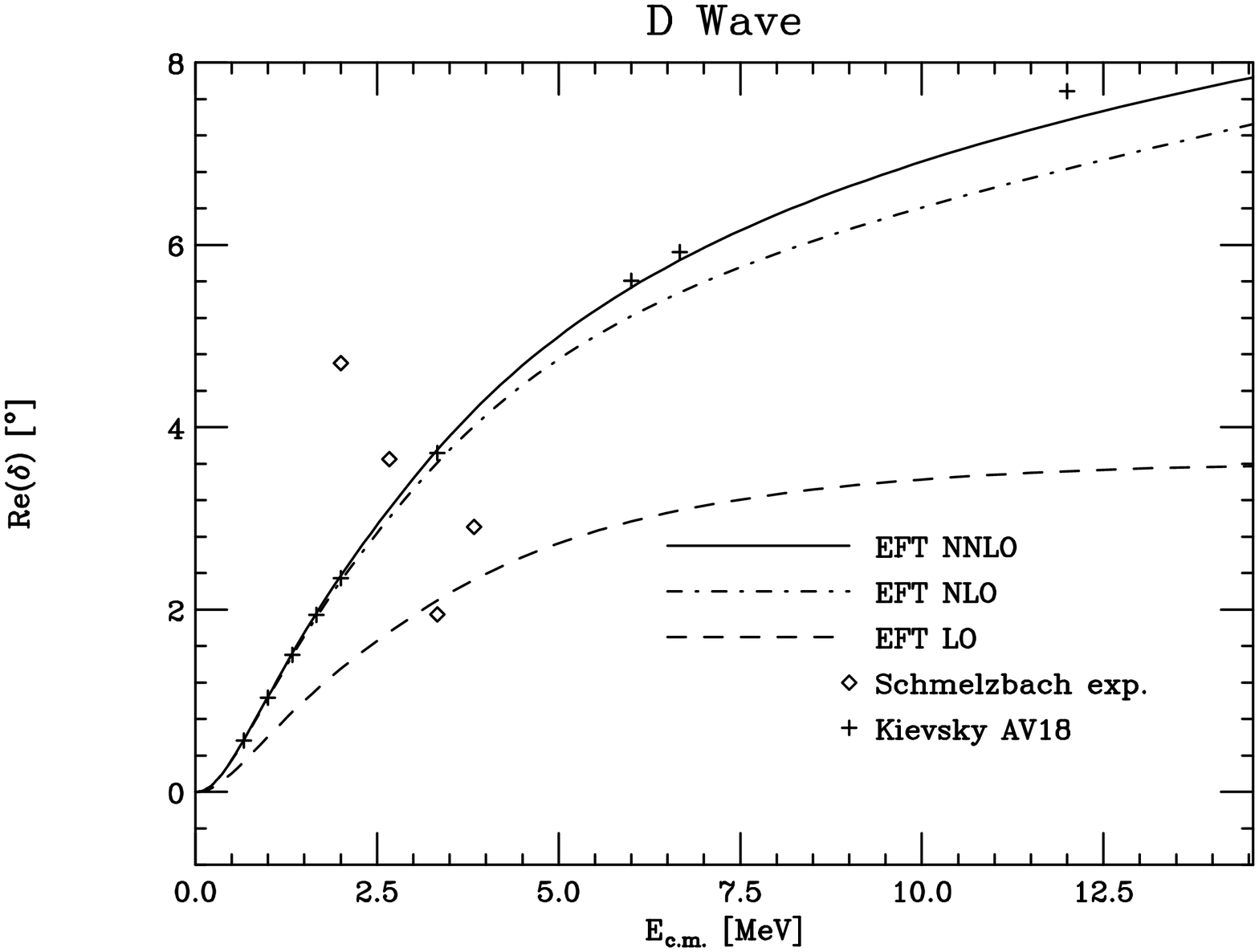} \hfill
      \includegraphics*[height=0.45\linewidth,angle=90]{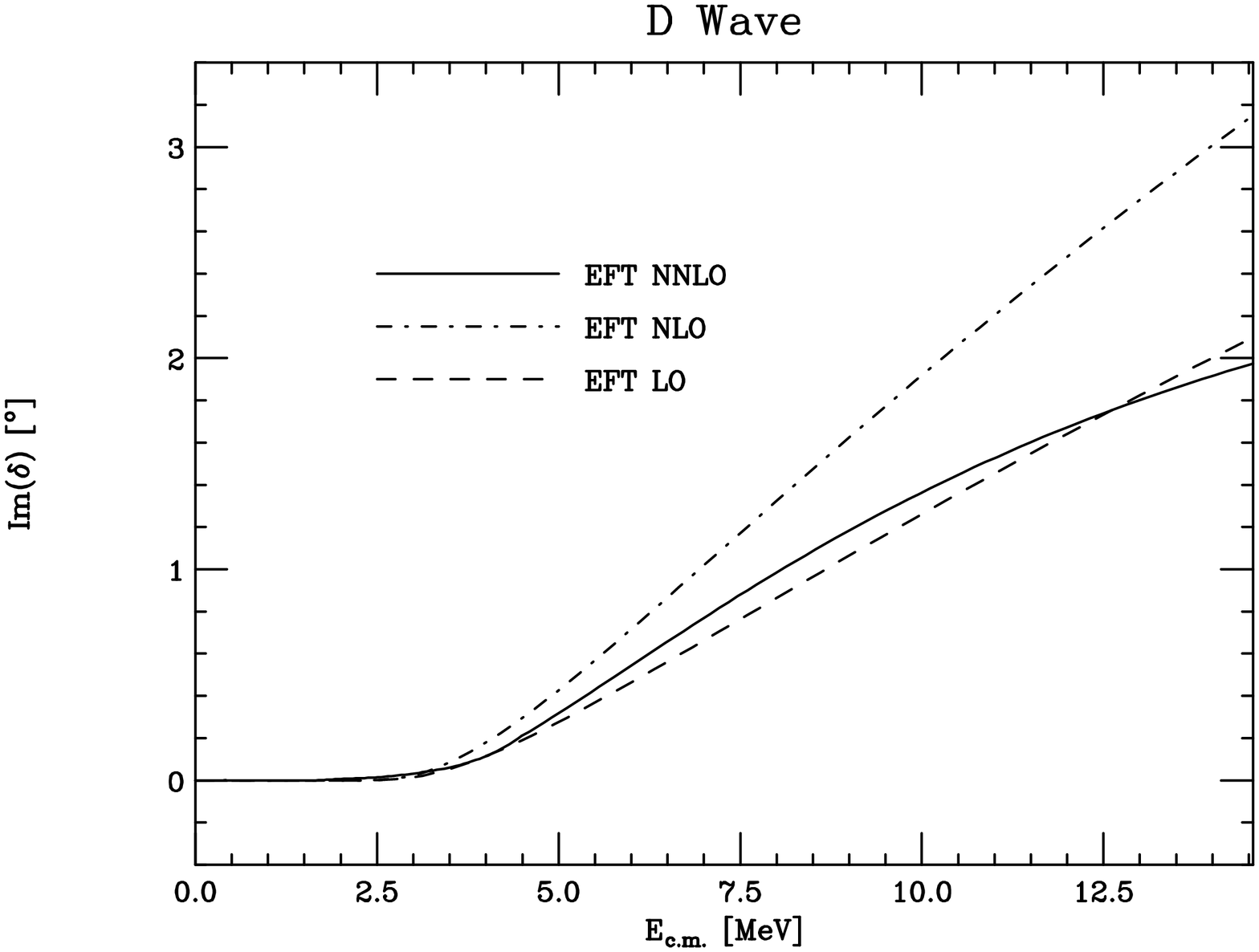} }
    
    \absatz \centerline{\includegraphics*[height=0.45\linewidth,
      angle=90]{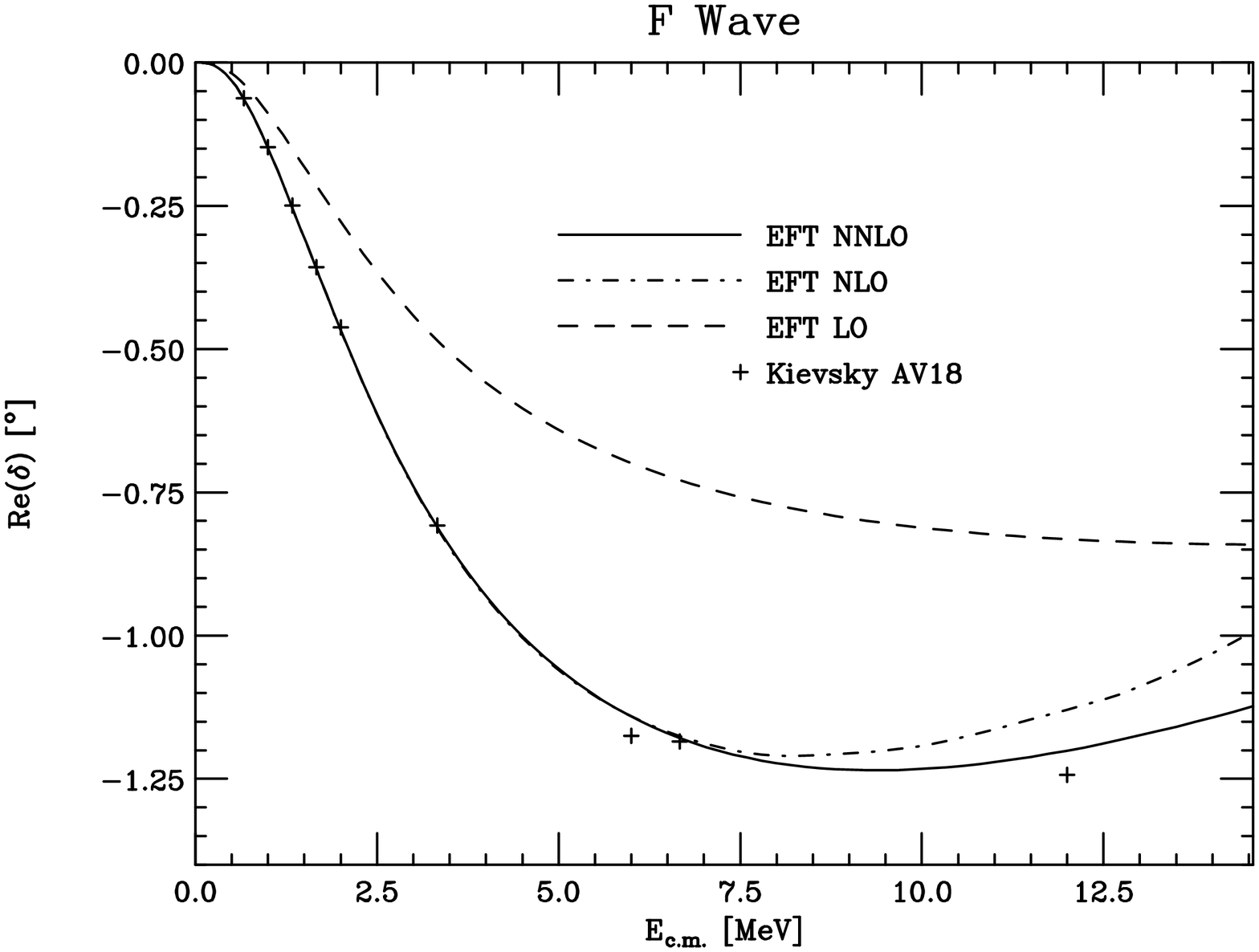} \hfill
      \includegraphics*[height=0.45\linewidth,angle=90]{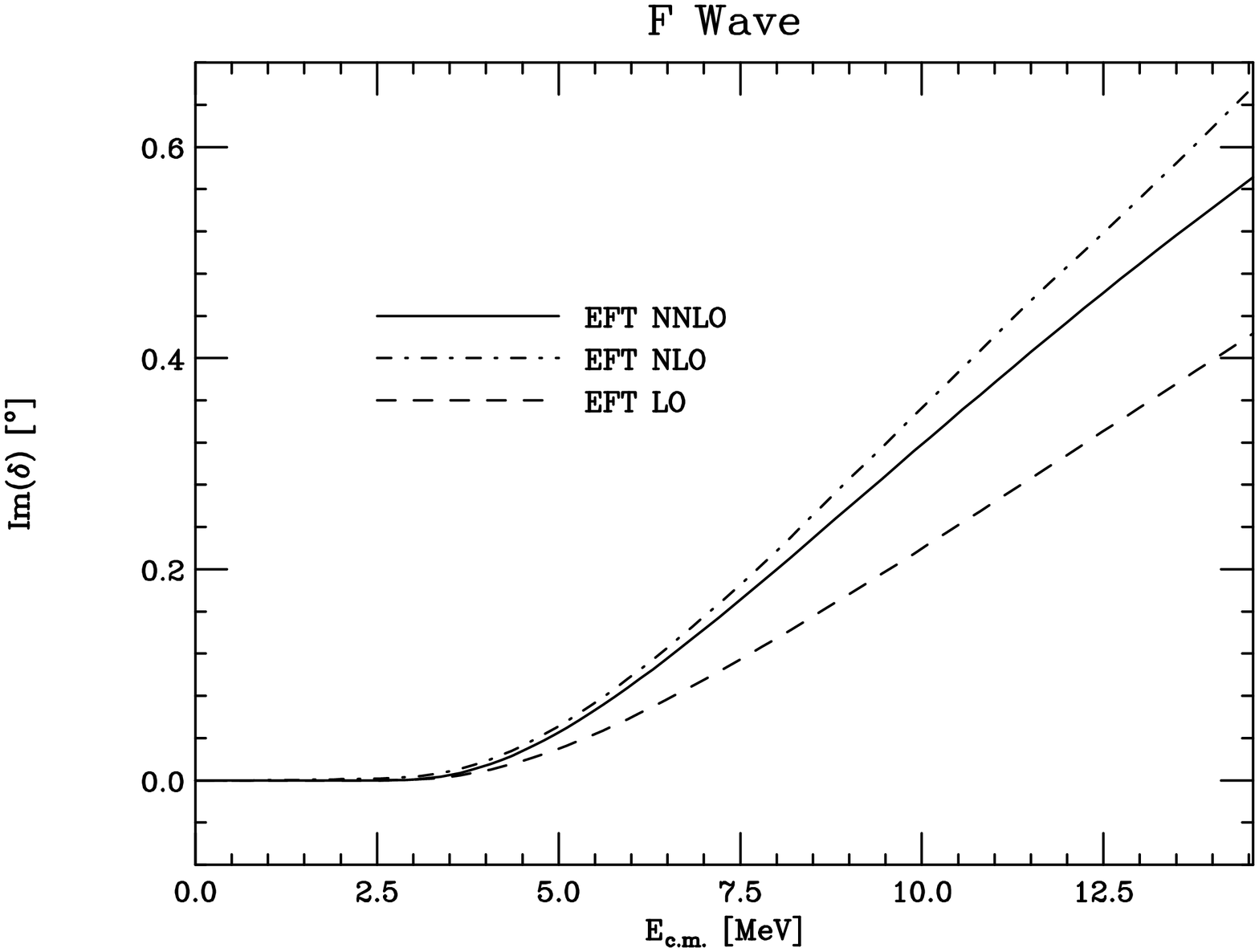} }
  
  \end{center}
\end{figure}

%%%%%%%%%%%%%%%%%%%%%%%%%%%%%%%%%%%%%%%%%%%%%%%%%%%%%%%%%%%%%%%%%%%%%%%%%%%%%%%
%%%%%%%%%%%%%%%%%%%%%%%%%%%%%%%%%%%%%%%%%%%%%%%%%%%%%%%%%%%%%%%%%%%%%%%%%%%%%%%
%%%%%%%%%%%%%%%%%%%%%%%%%%%%%%%%%%%%%%%%%%%%%%%%%%%%%%%%%%%%%%%%%%%%%%%%%%%%%%%

\section*{Acknowledgements}
We are indebted to G. Rupak, M.~J.~Savage and the effective field theory group
at the INT and the University of Washington in Seattle for a number of
valuable discussions, and to A.~Kievsky for communicating his results for the
higher partial waves above breakup prior to publication. F.~G.~thanks the
Nuclear Theory Group of the University of Washington, Seattle, for its kind
hospitality. The work was supported in part by the Department of Energy grants
DE-FG02-96ER40945 (F.G.) and DE-FG03-97ER41014 (H.W.G.), and by the
Bundesministerium f{\"u}r Bildung und Forschung (H.W.G.).

\newpage

%%%%%%%%%%%%%%%%%%%%%%%%%%%%%%%%%%%%%%%%%%%%%%%%%%%%%%%%%%%%%%%%%%%%%%%%%%%%%%%
%%%%%%%%%%%%%%%%%%%%%%%%%%%%%%%%%%%%%%%%%%%%%%%%%%%%%%%%%%%%%%%%%%%%%%%%%%%%%%%
%%%%%%%%%%%%%%%%%%%%%%%%%%%%%%%%%%%%%%%%%%%%%%%%%%%%%%%%%%%%%%%%%%%%%%%%%%%%%%%

%%%%%%%%%%%%%%%%%%%%%%%%%%%%%%%%%%%%%%%%%%%%%%%%%%%%%%%%%%%%%%%%%%%%%%%%%%%%%%%

\end{fmffile}
\end{document}